\documentclass[a4paper,oneside]{statsoc}

\usepackage[a4paper]{geometry}
\usepackage{graphicx}
\usepackage[textwidth=8em.textsize=small]{todonotes}
\usepackage{amsmath}
\usepackage{natbib}
\usepackage[flushleft]{threeparttable}
\usepackage{siunitx}
\usepackage{array,multirow,graphicx}
\usepackage[title]{appendix}
\usepackage{subfig}
\usepackage[paper=portrait,pagesize]{typearea}
\usepackage{makecell}
\usepackage{multirow}
\usepackage{adjustbox}
\usepackage{caption}

\graphicspath{ {./images/} }

\newcommand*{\showfontsize}{Fontsize: 1\@ptsize\,pt}

\title[Temperature-Driven non-linear Ecological Models]{Inference and model determination for Temperature-Driven non-linear Ecological Models}
\author[Author 1 {\it et al.}]{Marios Kondakis, Nikolaos Demiris and Ioannis Ntzoufras}
\address{Athens University of Economics and Business, Athens, Greece.}
\author[M. Kondakis et al.]{and Nikos E. Papanikolaou}
\address{Agricultural University of Athens, Athens, Greece.}
\address{Greek Ministry of Rural Development and Food, Athens, Greece.}

\begin{document}
\begin{abstract}
This paper is concerned with a contemporary Bayesian approach to the effect of temperature on developmental rates. We develop statistical methods using recent computational tools to model four commonly used ecological non-linear mathematical curves that describe arthropods' developmental rates. Such models address the effect of temperature fluctuations on the developmental rate of arthropods. In addition to the widely used Gaussian distributional assumption, we also explore Inverse Gamma--based alternatives, which naturally accommodate adaptive variance fluctuation with temperature. Moreover, to overcome the associated parameter indeterminacy in the case of no development, we suggest the Zero Inflated Inverse Gamma model. The ecological models are compared graphically via posterior predictive plots and quantitatively via Marginal likelihood estimates and Information criteria values. Inference is performed using the Stan software and we investigate the statistical and computational efficiency of its Hamiltonian Monte Carlo and Variational Inference methods. We explore model uncertainty and use Bayesian Model Averaging framework for robust estimation of the key ecological parameters.\\[1em]
\emph{keywords:} Arthropods' Developmental rates, Bayesian Model Averaging, Ecological models, Marginal likelihood estimation, Model comparison, Bayesian computation
\end{abstract}

\addtolength{\footnotesep}{3mm}
\footnotetext[0]{\textit{Address of correspondence:} Marios Kondakis, Department of Statistics, Athens University of Economics and Business, 76 Patission str., PO 10434, Athens, Greece. \\
e-mail:kondakis@gmail.com}

\section{Introduction}
Studying the population evolution of arthropod pests, as well as of biological control agents, is of great importance for the crop primary production, agricultural infrastructures, spreading diseases and consequently the economy \citep{bradshaw2016massive}. Temperature and body size are two major determinants that influence the metabolic, survival, growth and reproduction rates which control the ecological processes at all levels of arthropods' life \citep{brown2004toward}. Biological control is facilitated when the climate responses of biocontrol agents are understood, especially to temperature. The thermal thresholds for insect development can be estimated using several functional forms \citep{kontodimas2004}.

In all but the simplest cases, mathematical modelling is an indispensable tool for understanding the resulting developmental scheme \citep{kontodimas2004}.
However, fitting ecological models for developmental rates is not straightforward, typically because the mathematical forms are not linear \citep{papanikolaou2019elucidating} and the actual biochemical reactions of insects or  environmental factors responsible for their growth may remain unobservable. Therefore, we adopt the Bayesian paradigm to population dynamics' modeling and inference since it naturally accounts for latent parameters and their uncertainties. Nonetheless, there are significant challenges in designing statistical methods that work efficiently in a wide range of ecological applications. Stan \citep{carpenter2017stan} provides a BUGS-like interface to model building and the ability to run it via different languages and operating systems.

\subsection{Motivation for modeling developmental rate}
Insects and mites, as ectotherms, regulate their body temperature according the environment they live in \citep{norris2012effects}. This affects the rate of metabolism, i.e., the biochemical reactions that allow the processes of production and release of energy, as well as the synthesis of necessary molecules that serve as structural or functional components \citep{neven2000physiological}. In fact, temperature affects the functionality of enzymes, which act as catalysts for these biochemical reactions \citep{van2008slaves}. Consequently, within a range of temperatures in which insects and mites develop and reproduce, various biological features are affected \citep{broufas2001development,huey2001temperature,broufas2007development,nedvved2009temperature,papanikolaou2013temperature,papanikolaou2014life}. Thus, their performance is indebted to several temporal fluctuations in terms of population size through time. 
The empirical finding of the initial increase in the growth rate of insects and mites in relation to temperature, followed by its sharp decline, formed the basis for the development of various mathematical models of its description \citep{kontodimas2004}. These models allow the estimation of the lower and upper thermal limits, i.e. the lowest and highest temperature, respectively, at which the growth rate is zero, as well as the temperature at which it receives its maximum value. Understanding populations’ growth rate is of importance, as their assessment can lead to decisions on their management \citep{hare2011understanding}, particularly under the pressure of climatic change \citep{bradshaw2016massive}.

\subsection{Historical overview with models and techniques used in literature}

This paper investigates some popular non-linear ecological models that describe the rate of insects' and mites' development within a Bayesian context. We explore the computational and statistical efficiency of Hamiltonian Monte Carlo (HMC) \citep{betancourt2017conceptual,neal1901mcmc} and Variational Bayes (VB) \citep{blei2017variational}, a challenging task in the present setting due to distinct features of the entertained models, including truncation. Both the widely used Gaussian distributional assumption and a newly-developed Inverse Gamma-based version are explored in order to model the developmental rate distribution of insects and mites.  We compare the models using information criteria, models marginal likelihood estimates and graphical tools. In addition, model averaging techniques are used to provide robust estimates of the parameters of interest. A distinct feature of these models is that zero count data make one parameter of interest indeterminable and model fit potentially misleading. We propose ways to overcome this indeterminacy by applying the Zero Inflated Inverse Gamma distribution while carefully connecting the probability of non-zero development to the predictor.

The remainder of this paper is structured as follows. The next section contains the ecological models we develop. Section \ref{dataapplication} investigates two separate real-life examples without and with zero-rates, and Section \ref{results} reports our findings, while the paper concludes with discussion.

\section{Ecological models}
\label{Ecomod}
A variety of different linear and non-linear equations have been used to describe the rate of development of insects and mites and to estimate their thermal limits. Such linear approximations enable the calculation of lower developmental threshold and thermal constant within a small temperature range, usually 15-30\textsuperscript{o}C \citep{campbell,wagner,jarovsik2002,kontodimas2004}. However, the relationship between development and temperature becomes non-linear outside that range. Thus, in order to accurately predict developmental rates across the spectrum, the use of non-linear ecological models is required \citep{wagner,kontodimas2004,damos2012temperature}.

\subsection{Non-linear ecological models}
In the present work, four commonly used non-linear ecological models are considered. Specifically, the Bieri \citep{bieri1983development}, the Briere \citep{briere1998comparison,briere1998modeling,briere1999novel}, the Analytis \citep{analytis1981relationship} and the Lactin \citep{lactin1995improved} models are implemented. Developmental time is the duration between life stages of the insects or mites \citep{wagner}. The response variable $y(T;{\theta})$ describes the developmental rate and it is defined as the reciprocal of the days until the completeness of a particular developmental event. Herein, $T$ denotes the predictor variable, the absolute temperature measured in Celsius degrees, while ${\theta}$ denotes the parameter vector of the model. Typically, the expected developmental rate, $r(T;{\theta})$ is modelled and the four aforementioned models are presented below. 

\subsubsection{Bieri Model}
\begin{figure}[hbt!]
\begin{center}
\includegraphics[height=7.5cm]{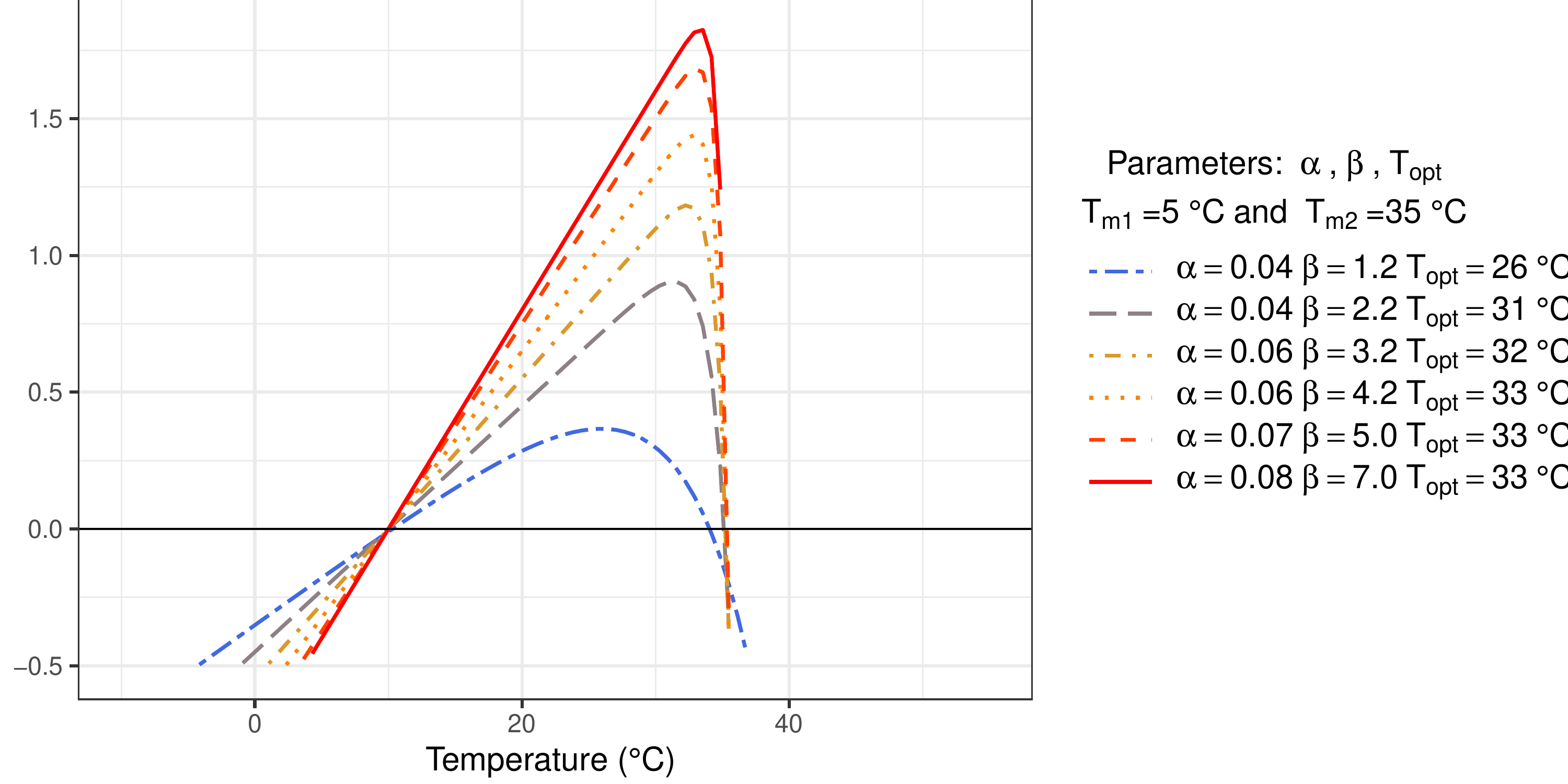}
    \caption{Bieri developmental rates}%
    \label{fig:Bierifun}%
    \end{center}
\end{figure}

In the Bieri model, the developmental rate is defined as
\begin{equation}\label{bieri_def}
    r(T;{\theta}) = \alpha\cdot(T-T_{m_1}) - {\beta}^{(T-T_{m_2})}
\end{equation}
where $\alpha$, $\beta$, $T_{m_1}$ and $T_{m_2}$ are the model parameters. In particular, the values of $T_{m_1}$ and $T_{m_2}$ lie close to the real lower and upper thermal thresholds $T_{min}$ and $T_{max}$, at which the development starts or ceases respectively. The exact values of the thermal thresholds are derived implicitly as the lower and upper roots respectively of the response function. The parameter $\alpha$ corresponds approximately to the rate of increase in the linear model at vital temperatures \citep{bieri1983development}. The response variable in (\ref{bieri_def}) is a concave function of the temperature. Parameter $\alpha$ is defined in the interval $\left(0,1 \right)$ while $\beta$ determines the decrease of the developmental rate at higher temperatures \citep{bieri1983development} when $\beta$ exceeds unity. According to (\ref{bieri_def}), we observe that: \\
\hspace*{\fill} $r(T_{m_1})= -{\beta}^{(T_{m_1}-T_{m_2})}$ and $r(T_{m_2})={\alpha}(T_{m_2}-T_{m_1}) -1$, \hspace*{\fill}\\
which suggests that $r(T_{m_1}), r(T_{m_2}) \in (-1,0)$ and leads to the following inequality
\begin{equation*}
    r(T_{m_1}) < r(T_{min}) < r(T_{max}) < r(T_{m_2}).
\end{equation*}     
The temperature at which the maximum developmental rate occurs is called optimum and it is denoted by $T_{opt}$. In the Bieri model, it is given by
\begin{equation*}\label{bieri_opt}
     T_{opt}=T_{max} + \frac{\log{\alpha}-\log\left ( {\log{\beta}} \right )}{\log{\beta}}.
    \end{equation*}
    \vspace{0.1cm}
In Fig. \ref{fig:Bierifun} some curves are generated by the Bieri model in (\ref{bieri_def}) when $T_{m_1}=5^{\circ}$C and $T_{m_2}=35 ^{\circ}$C while the other parameters vary. 

\subsubsection{Briere Model}
\begin{figure}[hbt!]
\begin{center}
\includegraphics[height=7.5cm]{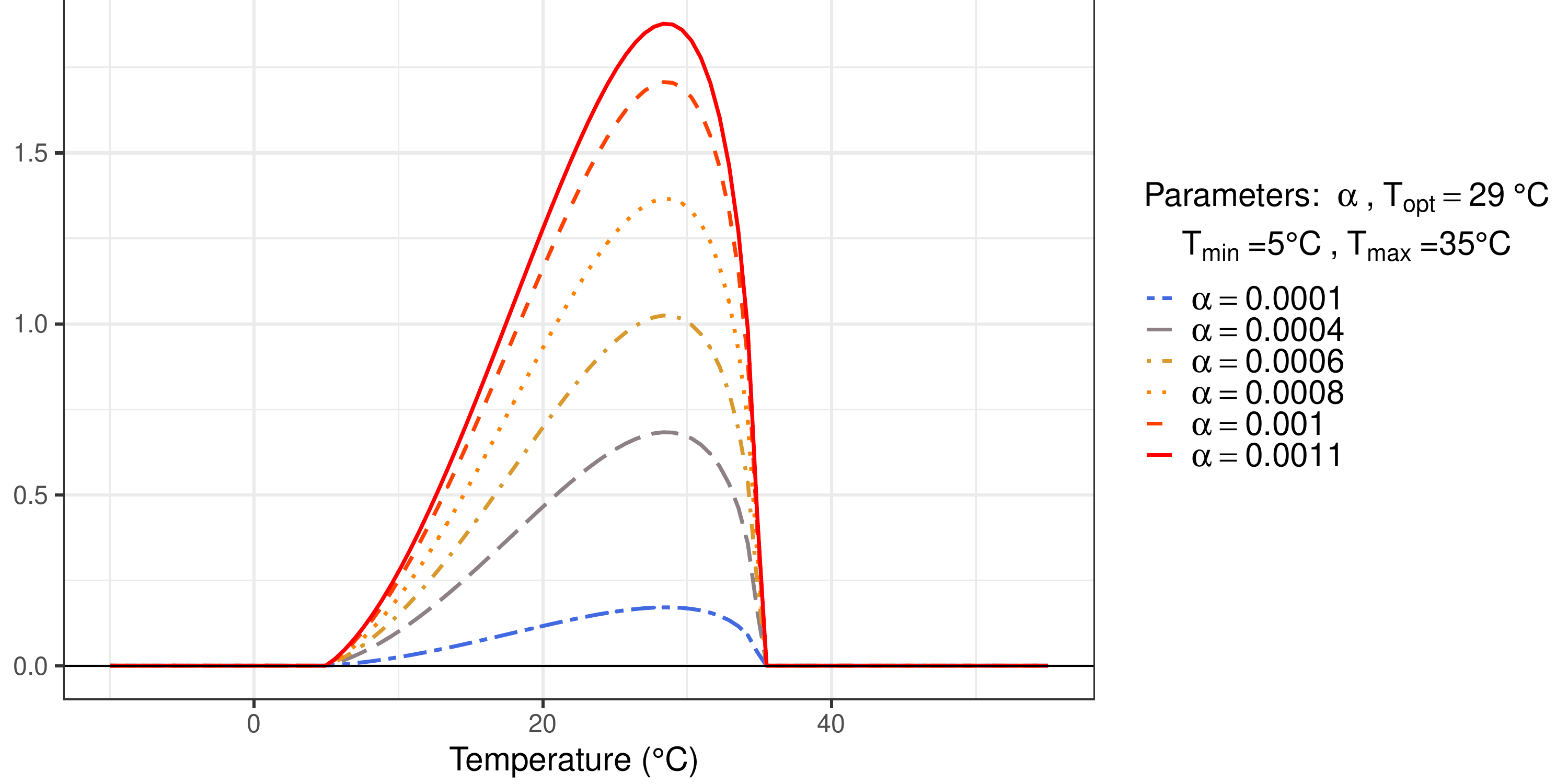}
    \caption{Briere developmental rates}%
    \label{fig:Brierefun}%
    \end{center}
\end{figure}

The Briere model is the most popular and parsimonious model. The developmental rate is defined as
\begin{equation}\label{briere_def}
    r(T;{\theta}) = \begin{cases} 
     \alpha \cdot T \cdot (T-T_{min}) \cdot \sqrt{(T_{max}-T)} & \mbox{ for }\, T_{min} < T < T_{max}\\
     0 & \mbox{ otherwise }\\
     \end{cases}
     \end{equation}
where $\alpha$, $T_{min}$ and $T_{max}$ are model parameters. Particularly, $T_{min}$ and $T_{max}$ are exactly the lower and upper thermal thresholds at which the development starts or ceases respectively while parameter $\alpha$ is an empirical constant \citep{briere1998modeling}. The response variable in (\ref{briere_def}) is again a concave function of the temperature. Parameter $\alpha$ is defined in $\left(0,1 \right)$ whereas the existence of the square root in (\ref{briere_def}) ensures that the developmental rate declines sharply at higher temperatures. The optimum temperature in the Briere model is given by
\begin{equation*}\label{briere_opt}
T_{opt} =\frac{1}{10}\left\{4\cdot T_{max}+3 \cdot T_{min}+\sqrt{{(4 \cdot T_{max}+3 \cdot T_{min})}^2-40 \cdot T_{min} \cdot T_{max}}\right\}.
\end{equation*}   
In Fig. \ref{fig:Brierefun} some curves are created by the Briere model in (\ref{briere_def}) when $T_{min}=5^{\circ}C$ and $T_{max}=35^{\circ}C$ while the parameter $\alpha$ varies.

\subsubsection{Analytis Model}
The developmental rate in Analytis model is defined by
\begin{equation}\label{analytis_def}
    r(T;{\theta}) = \begin{cases} 
     \alpha \cdot {(T-T_{min})}^n \cdot {(T_{max}-T)}^m & \mbox{ for }\, T_{min} < T < T_{max}\\
     0 & \mbox{ otherwise }\\
     \end{cases}
     \end{equation}
where $\alpha$, $T_{min}$, $T_{max}$, $n$ and $m$ are parameters of this model. The exponents $n$ and $m$ in (\ref{analytis_def}) are empirical constants that determine the rate of growth and decrease of the developmental rate respectively \citep{analytis1981relationship}. Both these parameter take values in $(0,+\infty)$ but in order to reduce the computational burden we may restrict them in a subset of the form $(0,c)$, for some constant $c>0$. Finally, $\alpha$ takes values in $\left(0,1 \right)$ interval. The optimum temperature in this model is given by
\begin{equation*}\label{analytis_opt}
T_{opt}=\frac{ n \cdot T_{max} + m \cdot T_{min}}{n+m}.
\end{equation*}

\begin{figure}[hbt!]
\begin{center}
\includegraphics[height=7.5cm]{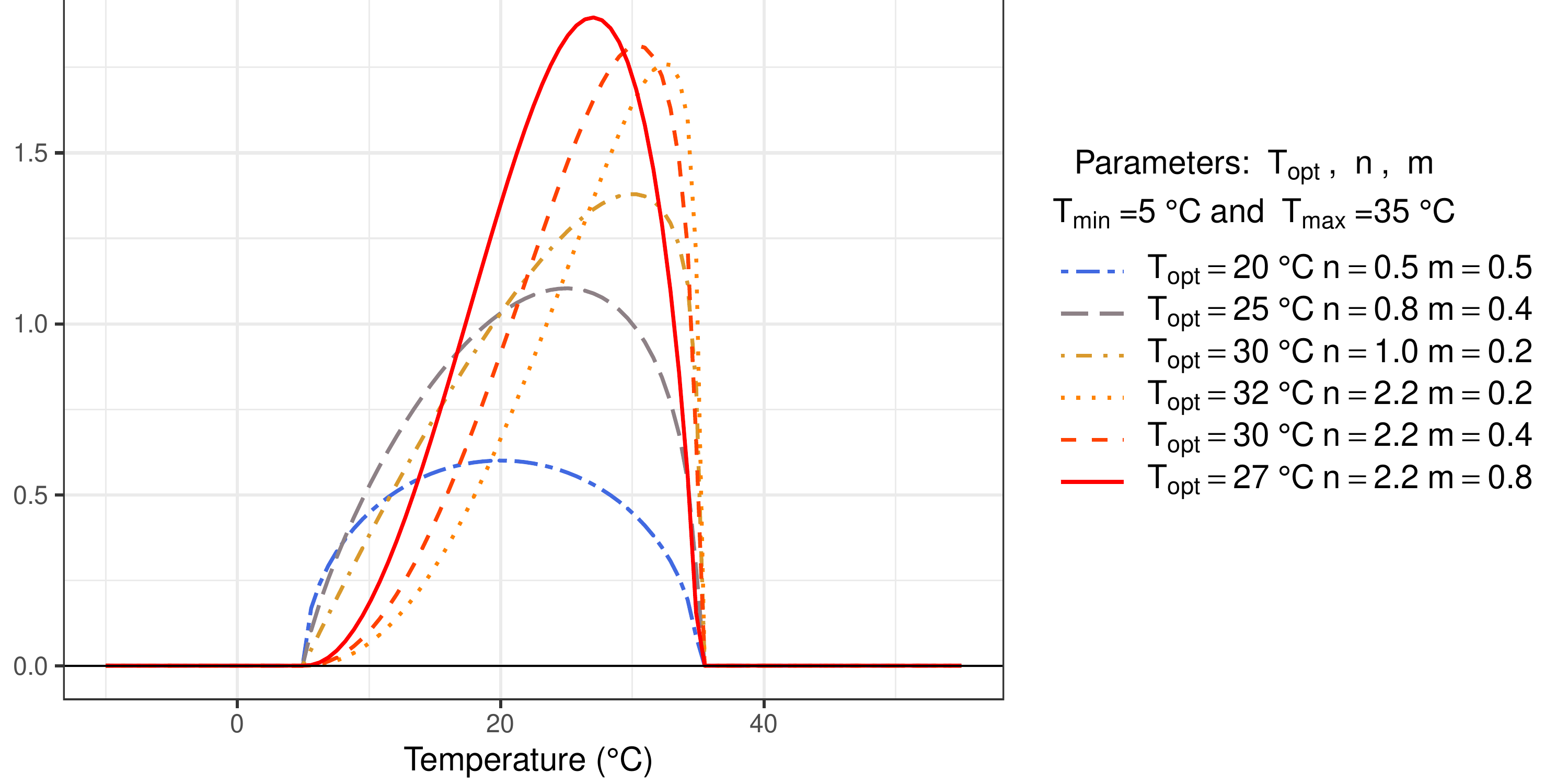}
    \caption{Analytis developmental rates}%
    \label{fig:Analytisfun}%
    \end{center}
\end{figure}

The Analytis model has a multiplicative polynomial structure in which the exponents change as parameters to be estimated. Such a structure needs some empirical driven tuning when defining its thermal parameters space. Especially, in the case of one dataset, we assumed that $T_{min}$ is greater than ${4^{\circ}C}$. Some curves generated by the Analytis model in (\ref{analytis_def}) are depicted in Fig. \ref{fig:Analytisfun} when $T_{min}=5^{\circ}C$ and $T_{max}=35^{\circ}C$, while parameters $n$ and $m$ vary.

\subsubsection{Lactin Model}
\begin{figure}[hbt!]
\begin{center}
\includegraphics[height=7.5cm]{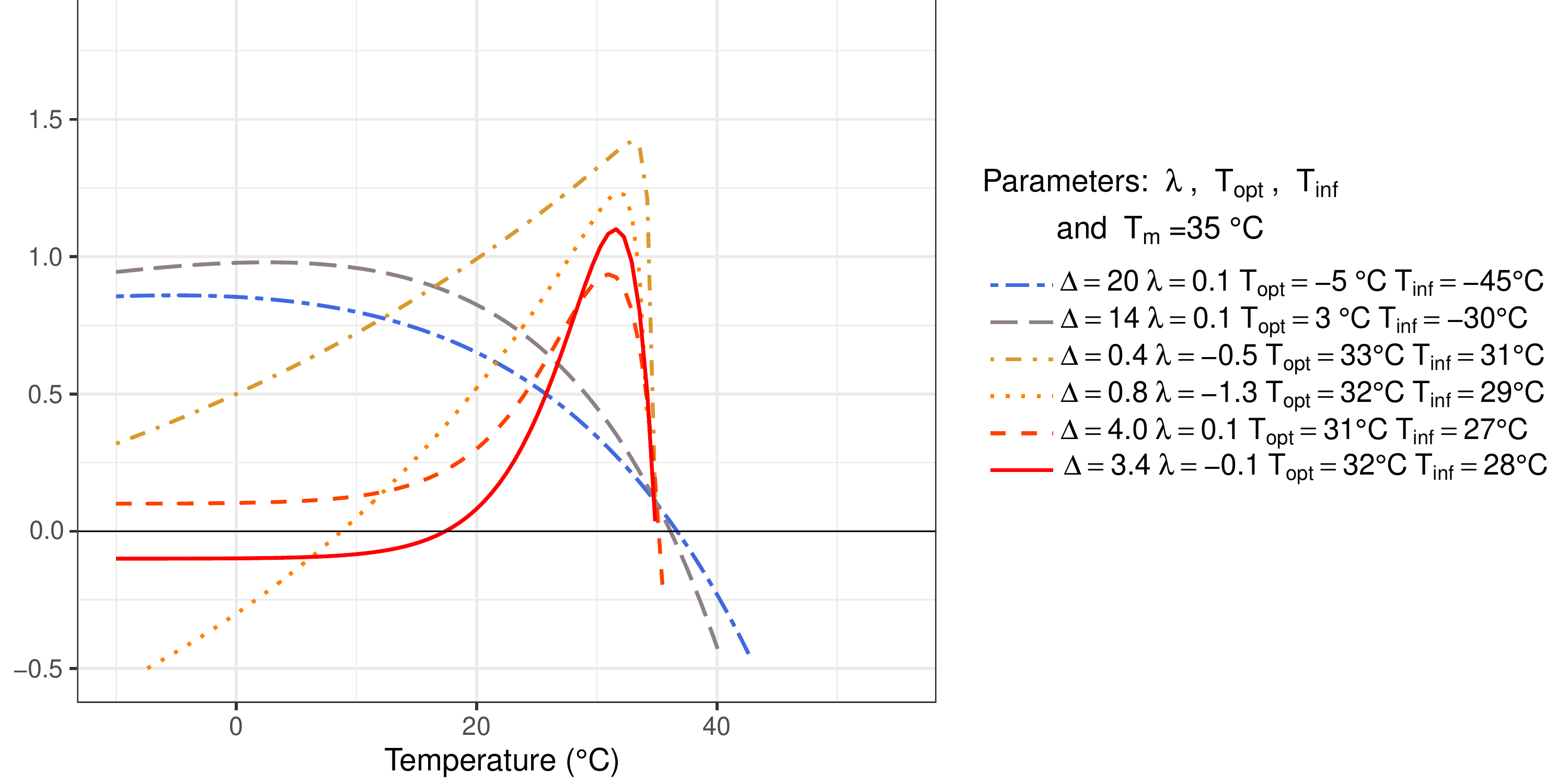}
    \caption{Lactin developmental rates}%
    \label{fig:Lactinfun}%
    \end{center}
\end{figure}

The Lactin model includes four parameters and the developmental rate is defined as
\begin{equation}\label{lactin_def}
r(T;{\theta})= \lambda+e^{\rho \cdot T}-e^{\rho \cdot T_m -\frac{(T_m-T)}{\varDelta}}
\end{equation}
where $T_m$ is associated with the upper thermal threshold $T_{max}$ since it tends to this value when $\lambda$ tends to be zero. Parameter $\lambda$ represents an asymptotic level of the developmental rate value in (\ref{lactin_def})  that is approximated when predictor $T$ tends either to $-\infty$ (extremely low temperatures) or to the threshold parameter $T_m$. Thus, in the event that the $\lambda$ is non-negative, the Lactin ecological function does not have a lower thermal threshold ($T_{min}$) and at the same time $T_{m} \leq T_{max}$ as the developmental rate value in (\ref{lactin_def}) is limited above zero level, at which the maximum thermal thresholds undergoes. In the case that $\lambda$ is negative, $T_{m} > T_{max}$ and $T_{min}$ sample space is in the $\left ( -\infty , T_{max} \right )$ interval. Parameter $\varDelta$ is positive and it determines the descent steepness of the developmental rate. It expresses the temperature range between the value at which the response function begins to descend and the value of the $T_m$ parameter. When $\varDelta$ is less than one, the rate of descent is very high, although in the other case the rate of descent is lower and similar to the other ecological models. This feature triggers discontinuity and lack of fit problems. Hence, in this work we define $\varDelta$ in $(1, +\infty)$ interval in order to avoid such problems. Parameter $\rho$ describes the acceleration of the function from low temperatures to the optimal temperature \citep{lactin1995improved}. 
The response function in (\ref{lactin_def}) has one inflection point, a maximum point at the optimum temperature and asymmetry about this point (left skewed). Also the function has a sharp drop after the optimum temperature, which is achieved by setting $\rho$ in $(0, {\varDelta}^{-1})$. The actual upper thermal threshold $T_{max}$ is evaluated as the higher root of response function in (\ref{lactin_def}). The optimum temperature in the Lactin model is given by
\begin{equation}\label{lactin_opt}
  T_{opt}= T_m-
    \frac{\varDelta \cdot \log{(\rho \cdot \varDelta)}}{\rho \cdot \varDelta - 1}.
\end{equation}
In addition, the temperature at the inflection point of the Lactin curve is given by
\begin{equation}\label{lactin_inf}
  T_{inf}= T_{opt}-\frac{\varDelta \cdot \log{(\rho \cdot \varDelta)}}{\rho \cdot \varDelta - 1}.
\end{equation}
In Fig. \ref{fig:Lactinfun} some curves are created by the Lactin model in (\ref{lactin_def}) when $T_{m}=35^{\circ}C$ and the parameters $\Delta$, $\rho$ and $\lambda$ vary.

\subsection{Ecological features of the models}
There are some basic common features in all of the above-mentioned ecological models. 

There is no growth below the lower temperature threshold  $T_{min}$ or above the upper temperature threshold $T_{max}$. Specifically, in the case of the Briere and Analytis models, the developmental rate is positive and is defined only between the two parameters of the thermal thresholds. Also, the developmental rate in the Bieri and Lactin models can take negative values that cannot be interpreted. To accommodate these characteristics, we use initial values of the parameters that allow the ecological function to receive positive values.

The developmental rate is an asymmetric curve left skewed of its maximum point. It increases and reaches a maximum at optimal temperature while it declines rapidly down to zero at the higher temperature threshold $T_{max}$ that is considered as lethal temperature.  It includes an inflection point, with the exception of the Bieri model which is a concave function of the temperature. The structure of the Lactin model makes it susceptible to a type of exponential pattern in data-values (Fig. \ref{fig:Lactinfun}). The Briere model, on the contrary, has a specific structure, which has the first derivative of class $\mathcal{O}\left(T^{2.5}\right)$ and can hardly trace exponential changes in data values (Fig. \ref{fig:Brierefun}). Also, the model of Analytis has polyonimic structure as the model of Briere but it does contain more degrees of freedom because the exponents it includes are unknown parameters (Fig. \ref{fig:Analytisfun}). The Bieri model also adopts an exponential reduction after the thermal optimum threshold but can only follow the developmental increase of the dataset linearly (Fig. \ref{fig:Bierifun}). All four ecological models have been used in the literature to provide reasonable estimates of the thermal thresholds of several anthropods' developmental rates at various stages \citep{bieri1983development,kontodimas2004,aghdam2011evaluation}. However, the Bieri model is underutilized in the literature, so we include it in this study to gain a better understanding of ecological models that describe temperature-dependent development.  Additionally, in \citep{kontodimas2004,aghdam2011evaluation} the ecological models were compared based on the accuracy of the real data thermal threshold estimates, the adjusted coefficients of regression ($R^2$), and the residual sum of squares values. In summary, the Briere and Analytis models, appear to overestimate and underestimate the upper and lower thermal thresholds, respectively, whereas the Bieri and Lactin models appear to meet the majority of the criteria used in the comparisons. Despite these minor differences, all of the above models appear to provide higher $R^2$ values than other models in real-data applications in the literature and we include all four in the current work.

\subsection{Measurement Error}
Probabilistic random error due to chance and systematic error due to data with excessive zeros is added in order to include uncertainty in the ecological models already listed. In section \ref{datalik}, we present the notation adopted and the probability schemes implemented in this analysis.

\subsubsection{The likelihood of the data} \label{datalik}
Let $y_i$ represent the observed developmental rates of the $i^{th}$ individual observed at $T_i$ temperature, where $i = 1,\dots,N$. We consider $y_i$ to be independent response variables counted as the reciprocal of the number of days until the development of the $i^{th}$ individual takes place and the range of its value lies in the $(0, 1)$ interval.
Furthermore, let $y$, $T$ and $\theta$ be the vectors of the response, the predictor and the parameters respectively then the conditional expectation $E(y|T,\theta)$ is considered to vary according to the ecological function $r(T;\theta)$ presented in (\ref{bieri_def}), (\ref{briere_def}), (\ref{analytis_def}) and (\ref{lactin_def}) respectively.\\
The data distribution is denoted by
\begin{equation*}\label{likel}
p({y}|{T,\theta})=\prod_{i=1}^{N} p(y_i|T_i,\theta).
\end{equation*}
In this study we consider the Gaussian and the Inverse-Gamma distributions as the distribution of the response data $y_i$. Specifically, the Gaussian distribution is used broadly in the literature as a good approximation to most unimodal distributions with finite variance due to the general form of the Central Limit Theorem. Thus, it can be used as to approximate a more complicated model likelihood of the data. The non-linear model of the independent response rates has the Gaussian distribution given by 
\begin{equation}\label{gausslike}
y_i \sim N(r(T_i;{\theta)},\sigma^2)
\end{equation}
where the mean of the Gaussian likelihood is driven by the respective ecological model, whereas its standard deviation $\sigma$ is considered as an unknown parameter. We consider a weakly informative prior distribution for $\sigma$ like: the Inverse-Gamma $ \text{Inv}\Gamma \left({10}^{-3},{10}^{-3}\right)$. 
The non-linear model used in (\ref{gausslike}) is a constant variance model among different temperatures. In addition, it allows for zero observed rate values, which occur when the stage of the insect does not change in perpetuity. Such modeling, however, has the disadvantage of lack of interpretability in the case of estimated negative rate values.

On the other hand, the Inverse-Gamma distribution is a plausible alternative for modeling positive observed rates, as it handles positive values that describe ratios such as developmental rates whose inverse are positive counts (such as days passed until the expected development occurs) that can be described by the Gamma distribution. The non-linear model of the independent response rates has the Inverse-Gamma distribution given by  
\begin{equation}\label{Igamma}
y_i \sim \text{Inv}\Gamma(\zeta, (\zeta-1)\cdot r({T_i};{\theta}))
\end{equation}
where $\zeta$ is the shape parameter of the Inverse-Gamma distribution. The mean of the distribution equals $r({T_i};{\theta})$, whereas the variance equals $\frac{r({T_i};{\theta})^{2}}{(\zeta-2)}$. 
The Inverse-Gamma likelihood in (\ref{Igamma}) is a natural alternative to model observed positive rates. Herein, its mean is driven by the respective ecological function, whereas its variance depends both to the shape parameter $\zeta$ and the ecological function $r({T_i};{\theta})$ as well, which allows the variance to be temperature dependent. For the prior distribution of the shape parameter $\zeta$ we have chosen a weakly informative $\Gamma \left({10}^{-1},{10}^{-2}\right)$.

\subsubsection{Zero-rates case}
There are situations in which any insect development does not occur throughout the cohort study. This is indicated by zero values in the response variable, which can theoretically be interpreted as the number of days required for the insect to move to its next stage never ending, implying that the developmental rate is the reciprocal of infinity. In such cases, the MCMC sampling procedure can be extended either to include prior information about the case of no development by adjusting the prior knowledge of the parameters concerned or to include a zero-inflation scheme.  
For this work, we suggest the use of a Zero Inflated Inverse Gamma distribution, which gives zero value with probability $p_{i}$ for the $i^{th}$ insect observed at $T_i$ temperature. Particularly, the probability density function of the observation $y_{i}$ is:
\begin{equation}
\begin{split}
P(y_i|T_i,\theta) =\left\{\begin{matrix} 
p_i \qquad  \qquad  \qquad  \qquad  \qquad \qquad \qquad \textrm{ if } y_i=0 
\\ P_{\text{Inv}\Gamma} \left \{ \zeta, (\zeta-1) \cdot r(T_i ;{\theta}) \right \} \cdot \left ( 1-p_i \right )   \textrm{ if } y_i\neq 0 
\end{matrix}\right. \\ 
\textrm{logit}\left (1-p_{i}\right )= \left [   c \cdot \left \{r(T_i ;{\theta}) -  k\right \} \right ] \Rightarrow p_i= \frac{1}{e^{  \left \{ c \cdot \left (r(T_{i};{\theta}) -  k\right ) \right \}   }+1} \label{pzero}
\end{split}    
\end{equation}


where $c$ is a constant positive parameter that has the opposite order of magnitude of the sample mean $\overline{y}$, while $k$ is the inflation point of logit link function where the probability of zero $p_{i}$ is equal to $\frac{1}{2}$. 
The constant $k$ is associated with the constant $c$ and can be chosen so that the zero rate of development matches the probability $p_i$ at a predetermined level like 0.9.
According to the real data example presented below in the results section, the proposed values that satisfy the above criteria for constants $c$ and $k$ are $10^2$ and $5 \cdot 10^{-3}$, respectively.

As the developmental rate of $r(T_i)$ increases to its maximum, the probability $p_i$ in (\ref{pzero}) decreases towards zero. On the other hand, when $r(T_i)$ tends to be a very small number, the probability $p_i$ in (\ref{pzero}) tends to be one.

\subsubsection{Priors}
In the Bieri model (\ref{bieri_def}) the prior distribution of the parameters $\alpha$, $\beta$, $T_{m_1}$ and $T_{m_2}$ are shown in Table \ref{tab:priors} respectively.

In the Briere model (\ref{briere_def}), the transformation $\widetilde{a}=-\log(\alpha)$  is considered instead of the original parameter $\alpha$. Weakly informative priors are considered for $\widetilde{a}$, $T_{min}$ and $T_{max}$ as shown in Table \ref{tab:priors} respectively. 
In the Analytis model (\ref{analytis_def}) the transformation $\widetilde{a}=-\log(\alpha)$ is considered  instead of $\alpha$ with the weakly prior distribution $\Gamma \left({10}^{-1},{10}^{-2}\right)$. 
Additionally, the prior distributions utilized for parameters $m$, $n$, $T_{min}$ and $T_{max}$ are given in Table \ref{tab:priors} respectively.

In the Lactin model (\ref{lactin_def}) a new parametrization is used. Hence, the new transformed parameters are:
\begin{equation*} \label{lactran1}
  l=-\lambda \quad, \quad del=\frac{1}{\varDelta} \quad, \quad  a={e}^{(\rho-del)\cdot T_{m}}
\end{equation*}    
are considered instead of the original parameters $\lambda$, $\varDelta$ and $T_{m}$, respectively. By their definition, the new parameters are taking values in $\lambda \in \mathcal{R}$, $0<{del}<1$ and $a \in \left(0, \rho\cdot\varDelta \right)$. The latter is derived by the definition of the optimum temperature in the Lactin model (\ref{lactin_opt}) and the fact that $T_{opt} > 0$. In addition, should the temperature at the inflection point $T_{inf}$ be a positive number, then from (\ref{lactin_inf}) we can get that $a \in \left(0, {\rho}^2\cdot{\varDelta}^2 \right)$. The prior distributions considered are shown in Table \ref{tab:priors} respectively for each ecological model.

\begin{table}

\begin{threeparttable}
\centering
\captionsetup{width=1\textwidth}
\caption{Priors of the parameters of the four ecological models}
\label{tab:priors}

\sisetup{table-number-alignment = center,round-mode = figures,
round-precision = 3}
\begin{tabular}{|c|c|c|c|c|c|}
\hline
Models                  & Parameters                                         & Priors                                          & Models                    & Parameters                                       & Priors                                    \\ \hline
\multirow{4}{*}{Bieri}  & $\alpha$                                           & U(0,1)                                          & \multirow{3}{*}{Briere}   & $^{\dag}$$\widetilde{a}=-\log(\alpha)$       & $\Gamma \left({10}^{-1},{10}^{-2}\right)$ \\ \cline{2-3} \cline{5-6} 
                        & $\beta$                                            & $\Gamma \left(2\cdot{10}^{-1},{10}^{-1}\right)$ &                           & $T_{min}$                                        & $\Gamma \left({10}^{-2},{10}^{-2}\right)$ \\ \cline{2-3} \cline{5-6} 
                        & $T_{m_1}$                                          & $\Gamma \left({10}^{-1},{10}^{-2}\right)$       &                           & $T_{max}$                                        & $\Gamma \left({10}^{-2},{10}^{-3}\right)$ \\ \cline{2-6} 
                        & $T_{m_2}$                                          & $\Gamma \left({10}^{-1},{10}^{-2}\right)$       & \multirow{5}{*}{Analytis} & $^{\dag}$$\widetilde{a}=-\log(\alpha)$ & $\Gamma \left({10}^{-1},{10}^{-2}\right)$ \\ \cline{1-3} \cline{5-6} 
\multirow{4}{*}{Lactin} & $^{\dag}$$l=-\lambda$                    & $\Gamma\left({10}^{-1},{10}^{-1}\right)$        &                           & $m$                                              & $\Gamma \left({10}^{-1},{10}^{-1}\right)$ \\ \cline{2-3} \cline{5-6} 
                        & $^{\dag}$$del=\frac{1}{\varDelta}$       & U(0,1)                                          &                           & $m$                                              & $\Gamma \left({10}^{-1},{10}^{-1}\right)$ \\ \cline{2-3} \cline{5-6} 
                        & $^{\dag}$$a={e}^{(\rho-del)\cdot T_{m}}$ & $\Gamma \left({10}^{-2},{10}^{-3}\right)$       &                           & $T_{min}$                                        & $\Gamma \left({10}^{-2},{10}^{-2}\right)$ \\ \cline{2-3} \cline{5-6} 
                        & $\rho$                                             & U(0,1)                                          &                           & $T_{max}$                                        & $\Gamma \left({10}^{-2},{10}^{-3}\right)$ \\ \hline
\end{tabular}

\begin{tablenotes}
\item{$\dag$} transformed parameter used.
\end{tablenotes}
\end{threeparttable}
\end{table}

\section{Application and results}
\label{dataapplication}
We adopt the Bayesian paradigm to inference and model selection. For concreteness we give the discernible features of the main tools we utilize in section  \ref{bayesian_details} in the appendix.
\subsection{Data application}
The reciprocals of the days counted express the observed rates of the insects and mites from egg to adult stage. The range of the observed rates are at [0, 1] interval. The zero value indicates that no development is observed. This situation implies the existence of a truncation point which gives explicitly an upper bound for the upper thermal threshold. 

Two datasets are used in this analysis. They concern the study of the two-spotted mite, \textit{Tetranychus urticae} \citep{barber2003biocontrol} and the fourteen-spotted ladybird beetle, \textit{Propylea quatuordecimpunctata} \citep{papanikolaou2013temperature}. The \textit{Tetranychus urticae} data developmental rates have minimum 0.019 at $15^{o}C$ and maximum 0.182 at $32.5^{o}C$. The \textit{Propylea quatuordecimpunctata} dataset consist of 105 beetles and their developmental rate  have minimum 0 at $35^{o}C$ and maximum 0.111 at  $32.5^{o}C$.

\subsection{\textit{Tetranychus urticae}}

There are 247 mites that have inserted adult stage until the study ended. The four ecological models are used both assuming the Gaussian and the Inverse Gamma distributions for the data. The information criteria along with the estimates of the marginal likelihood for each model are provided in Table \ref{tab:perakis_tetra}. 

\begin{table}
\resizebox{\linewidth}{!}{%

\begin{threeparttable}
\centering
\captionsetup{width=2\textwidth}
\caption{Model selection criteria for the eight models applied to the \textit{Tetranychus urticae} data}
\label{tab:perakis_tetra}

\sisetup{table-number-alignment = center,round-mode = figures,
round-precision = 3}
\begin{tabular}{|c|c|c|c|c|c|c|c|c|c|} 
\hline
                                                                                       &          & AIC               & DIC               & LooCV             & WAIC              & BIC               & $log\left(P_{y}IS\right)^{\dag}$ \,(se)& $log\left(P_{y}PP\right)^{\ddag}$ \,(se)& $log\left(P_{y}BS\right)^{\S}$ \,(se)\\ 
\hline
\multirow{4}{*}{\begin{tabular}[c]{@{}c@{}}Gaussian\\model \end{tabular}}      & Bieri    & -1663.2           & -1667.7           & -1672.0           & -1672.1           & -1638.6           & 796.1 (28.6)              & 777.2 (19.0)               & 792.6 (7.1)                 \\ 
\cline{2-10}
                                                                                       & Briere   & -1592.2           & -1595.8           & -1595.8           & -1595.9           & -1574.7           & 759.6 (39.0)              & 728.2 (12.4)               & 758.0 (6.9)                 \\ 
\cline{2-10}
                                                                                       & Analytis & -1738.3           & -1736.9           & -1744.0           & -1744.0           & -1717.3           & 815.9 (40.5)              & 830.2 (11.0)               & 821.6 (12.3)                \\ 
\cline{2-10}
                                                                                       & Lactin   & -1673.9           & -1675.6           & -1689.1           & -1689.1           & -1638.8           & 797.8 (40.0)              & 821.8 (23.2)               & 793.8 (56.4)                \\ 
\hline
\multirow{4}{*}{\begin{tabular}[c]{@{}c@{}}Inverse\\Gamma\\model \end{tabular}} & Bieri    & -1715.6           & -1716.0           & -1717.2           & -1717.2           & -1698.1           & 837.4 (42.6)              & 839.7 (9.0)                & 832.1 (10.0)                \\ 
\cline{2-10}
                                                                                       & Briere   & -1749.4           & -1749.4           & -1746.3           & -1746.3           & -1735.4           & 846.7 (46.1)              & 874.0 (12.8)               & 848.0 (3.5)                 \\ 
\cline{2-10}
                                                                                       & Analytis & -1889.0           & -1887.3           & -1889.0           & -1889.1           & -1867.6           & 904.2 (42.6)              & 928.2 (16.3)               & 903.9 (26.2)                \\ 
\cline{2-10}
                                                                                       & Lactin   & \textbf{-1911.4}  & \textbf{-1911.0}  & \textbf{-1910.5}  & \textbf{-1910.6}  & \textbf{-1893.9}  & \textbf{920.9 (30.3)}     & \textbf{956.4 (17.0)}      & \textbf{922.0 (6.5)}        \\
\hline
\end{tabular}

\begin{tablenotes}
\item{$\dag$} $log\left(P_{y}IS\right)$ denotes the logarithm of estimated marginal likelihood via Importance sampling,
\item{$\ddag$} $log\left(P_{y}PP\right)$ denotes the logarithm of estimated marginal likelihood via Power posterior,
\item{$\S$} $log\left(P_{y}BS\right)$ denotes the logarithm of estimated marginal likelihood via Bridge sampling.
\end{tablenotes}
\end{threeparttable}}
\end{table}

\begin{figure}
\begin{center}
\includegraphics[height=6cm,width=\linewidth]{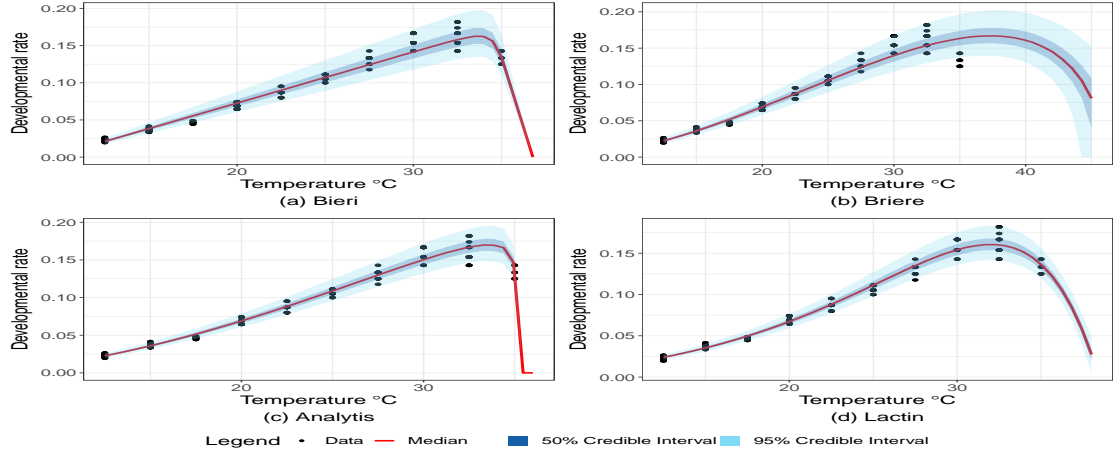}
    \caption{Predicted posteriors versus \textit{Tetranychus urticae} data using Inverse Gamma distribution}%
    
\label{fig:MFL4}%
    \end{center}
\end{figure}

Both Information criteria and marginal likelihood results clearly suggest that the Inverse Gamma distribution has better fit than the Gaussian distribution at \textit{Tetranychus urticae} dataset across all the ecological models. Furthermore the Lactin model with the Inverse Gamma distribution stands out against all the other cases. In the Gaussian case, the Analytis model excels.The use of the Inverse Gamma distribution, on the other hand, not only increases the Analytis model's efficiency but also adds flexibility to the Lactin model. The Briere model has the poorest criteria values (Table \ref{tab:perakis_tetra}). Although the
information criteria have fairly indicated the Briere model as the most suitable for the data, when we concentrate on the upper thermal threshold $T_{max}$ estimate, its performance is poor compared to the other ecological models shown in Fig. \ref{fig:MFL4}. Nevertheless, even though there is a clear picture concerning information criteria values between the ecological models, there is some variation between marginal likelihood estimates within ecological models in Table \ref{tab:perakis_tetra}.

Posterior means, $95\%$ credible limits and the effective sample size (neff) of the thermal thresholds and the deviance of the four ecological models are summarised in Tables \ref{tab:propyleahmc4} and \ref{tab:propyleahmc3} using the Inverse Gamma and the the Gaussian distribution, respectively. In addition, the HMC, ADVI meanfield and ADVI fullrank methods appear alternately in each column for each parameter of interest.

The $T_{min}$ credible limits estimates between the four ecological models do not overlap in Tables \ref{tab:propyleahmc4} and \ref{tab:propyleahmc3}. Bieri model has greater limits and Lactin model gives negative value estimates in the Inverse Gamma case Table \ref{tab:propyleahmc4}. The credible limits for $T_{opt}$ overlap between Bieri and Analytis whereas the estimates are lower for Lactin model and greater for Briere model. The $T_{max}$ credible limits overlap for Bieri and Lactin in Table \ref{tab:propyleahmc3}, has higher values for Briere and lower for Analytis model. 

 The credible limits using ADVI meanfield and fullrank methods overlap with the HMC credible limits in most cases in Tables \ref{tab:propyleahmc4} and \ref{tab:propyleahmc3}. Also the ADVI fullrank method seems to be closer to the HMC estimates as in Briere and Lactin models in Table \ref{tab:propyleahmc4}. However in general it has worst fit than the corresponding fitted model using HMC and also gives wider $95\%$ credible intervals. 
 
Bayesian model averaging provides alternative estimates for the parameters of interest, combining the predictive efficiency of all four ecological models. The derived weights are shown in Table \ref{tab:tetraw} whereas the BMA estimates and their $95\%$ credible intervals are given on Table \ref{tab:bma_tetra} and are divided into the data case of the Gaussian distribution and the data case of the Inverse Gamma distribution. The predictive bias of the Analytis and Lactin models appear to affect the model averaging estimates of $T_{min}$, $T_{max}$ and the deviance in the Gaussian and the Inverse Gamma case respectively. Also different weights based on (\ref{IC_weights}) give almost identical $95\%$ credible limits. However using ELBO based BMA weights do not give robust estimates of the parameters of interest which is not unexpected since the ELBO is a lower bound estimate of the marginal likelihood. 
Furthermore, the time elapsed until the completion of the algorithm for ADVI methods is up to 52 seconds, while for the HMC method is at least 476 seconds for Tetranychus dataset as shown in Table \ref{tab:etime1}. The average difference between the working times of HMC and each ADVI method is around 3500 seconds (58 minutes) whereas between the meanfiled and fullrank is around 18 seconds. These findings illuminate the VBI fullrank method's time-efficacy.

\begin{center}
\begin{adjustbox}{{angle=0},{scale=0.6}}
\begin{threeparttable}
\caption{Posterior summaries for the four models using the Inverse Gamma distribution for the \textit{Tetranychus urticae} data. In each column we report the HMC, the ADVI-meanfield and ADVI-fullrank estimates respectively.}
\label{tab:propyleahmc4}
\begin{tabular}{|c|c|c|c|c|c|c|c|c|c|} 
\hline
                                                                                &          &  \multicolumn{3}{c|}{$T_{min}$}& neff$^{\ddag}$                                                                                                                 &  \multicolumn{3}{c|}{$T_{opt}$}& neff$^{\ddag}$                                                                                                                               \\ 
\hline
\multirow{4}{*}{\begin{tabular}[c]{@{}c@{}}Mean\\ \& $95\%$\\ Cr. I.\end{tabular}} & Bieri    & \begin{tabular}[c]{@{}c@{}}9.4\\ (9.3, 9.6)\end{tabular}       & \begin{tabular}[c]{@{}c@{}}9.3\\ (9.1, 9.4)\end{tabular}       & \begin{tabular}[c]{@{}c@{}}9.3\\ (9.0, 9.5)\end{tabular}       & 19741 & \begin{tabular}[c]{@{}c@{}}33.7\\ (33.2, 34.1)\end{tabular}          & \begin{tabular}[c]{@{}c@{}}145.5\\ (134.8, 149.8)\end{tabular}       & \begin{tabular}[c]{@{}c@{}}203.2\\ (114.5, 326.6)\end{tabular}       & 8888   \\ 
\cline{2-10}
                                                                                & Briere   & \begin{tabular}[c]{@{}c@{}}6.6\\ (6, 7)\end{tabular}           & \begin{tabular}[c]{@{}c@{}}6.6\\ (6.4, 6.7)\end{tabular}       & \begin{tabular}[c]{@{}c@{}}6.5\\ (6.1, 7)\end{tabular}         & 7933  & \begin{tabular}[c]{@{}c@{}}36.7\\ (35.2, 38.5)\end{tabular}          & \begin{tabular}[c]{@{}c@{}}36.6\\ (36.1, 37)\end{tabular}            & \begin{tabular}[c]{@{}c@{}}36.6\\ (35, 38.3)\end{tabular}            & 6978   \\ 
\cline{2-10}
                                                                                & Analytis & \begin{tabular}[c]{@{}c@{}}4.2\\ (4, 4.6)\end{tabular}         & \begin{tabular}[c]{@{}c@{}}7.6\\ (7.5, 7.8)\end{tabular}       & \begin{tabular}[c]{@{}c@{}}6.0\\ (4.4, 9.4)\end{tabular}       & 9977  & \begin{tabular}[c]{@{}c@{}}33.6\\ (33.3, 34)\end{tabular}            & \begin{tabular}[c]{@{}c@{}}99.9\\ (96.6, 103)\end{tabular}           & \begin{tabular}[c]{@{}c@{}}79.9\\ (16.8, 239.3)\end{tabular}         & 7876   \\ 
\cline{2-10}
                                                                                & Lactin   & \begin{tabular}[c]{@{}c@{}}-18.7\\ (-18.7, -18.7)\end{tabular} & \begin{tabular}[c]{@{}c@{}}7.8\\ (7.6, 8.1)\end{tabular}       & \begin{tabular}[c]{@{}c@{}}-18.7\\ (-18.8, -18.7)\end{tabular} & 18393 & \begin{tabular}[c]{@{}c@{}}32.0\\ (31.8, 32.2)\end{tabular}          & \begin{tabular}[c]{@{}c@{}}32.1\\ (31.8, 32.3)\end{tabular}          & \begin{tabular}[c]{@{}c@{}}33.9\\ (33.8, 34.1)\end{tabular}          & 12228  \\ 
\hline
                                                                                &          & 
                                                                                
                          \multicolumn{3}{c|}{$T_{max}$}& neff$^{\ddag}$                                                                                                                                                 &
                          \multicolumn{3}{c|}{dev$^{\dag}$}& neff$^{\ddag}$                                                                                                                 \\ 
\hline
\multirow{4}{*}{\begin{tabular}[c]{@{}c@{}}Mean\\ \& $95\%$\\ Cr. I.\end{tabular}} & Bieri    & \begin{tabular}[c]{@{}c@{}}35.8\\ (35.4, 36.5)\end{tabular}    & \begin{tabular}[c]{@{}c@{}}150.8\\ (149.2, 152.5)\end{tabular} & \begin{tabular}[c]{@{}c@{}}209.0\\ (114.5, 326.6)\end{tabular} & 7738  & \begin{tabular}[c]{@{}c@{}}-1720.9\\ (-1724.9, -1713)\end{tabular}   & \begin{tabular}[c]{@{}c@{}}-1652.9\\ (-1658.6, -1637.6)\end{tabular} & \begin{tabular}[c]{@{}c@{}}-1652.4\\ (-1658.5, -16)\end{tabular}     & 10228  \\ 
\cline{2-10}
                                                                                & Briere   & \begin{tabular}[c]{@{}c@{}}45.0\\ (43.1, 47.3)\end{tabular}    & \begin{tabular}[c]{@{}c@{}}44.8\\ (44.2, 45.4)\end{tabular}    & \begin{tabular}[c]{@{}c@{}}44.9\\ (42.8, 47)\end{tabular}      & 6935  & \begin{tabular}[c]{@{}c@{}}-1753.3\\ (-1757, -1745.9)\end{tabular}   & \begin{tabular}[c]{@{}c@{}}-1752.9\\ (-1757.1, -1741.4)\end{tabular} & \begin{tabular}[c]{@{}c@{}}-1751.9\\ (-1756.7, -1740.4)\end{tabular} & 8638   \\ 
\cline{2-10}
                                                                                & Analytis & \begin{tabular}[c]{@{}c@{}}35.0\\ (35, 35.1)\end{tabular}      & \begin{tabular}[c]{@{}c@{}}99.9\\ (96.6, 103.1)\end{tabular}   & \begin{tabular}[c]{@{}c@{}}80.9\\ (17.5, 240.6)\end{tabular}   & 9038  & \begin{tabular}[c]{@{}c@{}}-1894.0\\ (-1899.2, -1885)\end{tabular}   & \begin{tabular}[c]{@{}c@{}}-1670.9\\ (-1691.8, -1641.4)\end{tabular} & \begin{tabular}[c]{@{}c@{}}-1495.1\\ (-1687.4, -149.5)\end{tabular}  & 12370  \\ 
\cline{2-10}
                                                                                & Lactin   & \begin{tabular}[c]{@{}c@{}}38.4\\ (38.1, 38.9)\end{tabular}    & \begin{tabular}[c]{@{}c@{}}42.6\\ (42.4, 42.8)\end{tabular}    & \begin{tabular}[c]{@{}c@{}}38.6\\ (38.2, 39)\end{tabular}      & 8284  & \begin{tabular}[c]{@{}c@{}}-1916.3\\ (-1920.6, -1908.3)\end{tabular} & \begin{tabular}[c]{@{}c@{}}-1705.8\\ (-1747.8, -1650.5)\end{tabular} & \begin{tabular}[c]{@{}c@{}}-1890.0\\ (-1918.5, -1791.8)\end{tabular} & 9938   \\
\hline
\end{tabular}
\begin{tablenotes}
\item{$\dag$} deviance of the model given the data.
\item{$\ddag$} effective sample size.
\end{tablenotes}
\end{threeparttable}
\end{adjustbox}
\vspace{0.5cm}

\end{center}
In Fig. \ref{fig:MFL4} the predicted posteriors versus \textit{Tetranychus urticae} data using Inverse Gamma distribution are shown for the ecological models.

The adaptivity in data across predictor values is clear in all the ecological models, where the $95\%$ credible limits are adjusted to data variance in each temperature level. On the other hand, in the Gaussian case (Fig. \ref{fig:MFL3}) the variance of the posterior remains constant across predictor values.
\subsection{\textit{Propylea quatuordecimpunctata} dataset}
There are 17 out of 105 insects that have not altered their egg status until the study ended. These cases are observed at ${35^{\circ}C}$ and are indicated by zeros in the response variable y. The Inverse Gamma distribution is not defined for zero response values. On the other hand, the Bieri and Lactin models can also generate negative values. Hence, in order to account for the presence of zeros, we use a Zero Inflated Inverse Gamma model and choose parameter initial values so as the scale of the Inverse Gamma to remain positive. Both the Gaussian and the Zero Inflated Inverse Gamma distributions are used for the data along with the four ecological models. The various model selection tools are summarised in Table \ref{tab:perakis}.

Both Information criteria and marginal likelihood results clearly suggest that the Zero Inflated Inverse Gamma distribution has better fit than the Gaussian distribution at Propylea quatordicempuncata only at the Analytis model. Furthermore Bieri and Analytis models stand out in the Gaussian and the Inverse Gamma case respectively, whilst the Briere model has the poorest criteria values (Table \ref{tab:perakis}). Nevertheless, even though there is a clear picture of the goodness of fit between the ecological models, there is some variation between marginal likelihood estimates within ecological models in Table \ref{tab:perakis}. 

In the Gaussian model, the Bieri and Lactin models stand out according to the information criteria and marginal likelihood estimates, the Analytis model follows, whilst the Briere model has the lower criteria values. On the contrary, in the Zero Inflated Inverse Gaussian model, the Analytis model is a better choice according to the marginal likelihood and the information criteria values, while Bieri, Lactin and Briere models follow respectively. Lactin and Bieri can be interpreted by their ability to track an exponential data-value decrease. Additionally, the Bieri model manages to capture the linear increase in the \textit{Propylea quatuordecimpunctata} dataset. The Briere model, on the other hand, lacks performance due to its unique structure, which requires that the decline of the response variable as the temperature rises be of the form $\sqrt{T_{max}-T}$. The Analytis model has similar multiplicative structure to the Briere model, but it is more complex model since the exponents of its model are unknown variables that make it adaptive and liable to data change especially in the Inverse Gamma case. Nevertheless, the Analytis model needs some tuning when defining its thermal parameters space. This is necessary to avoid allowing $T_{min}$ values close to zero, which would result in parameter underestimation and poor fit in the Analytis model. Especially, in the case of the Propylea dataset, we a-prior assumed that $T_{min}$ is greater than ${4^{\circ}C}$.
Posterior means, $95\%$ credible limits and the effective sample size of the thermal thresholds and the deviance of the four ecological models are summarised in Tables \ref{tab:propyleahmc2} and \ref{tab:propyleahmc1} using the Zero inflated Inverse Gamma and the Gaussian distribution respectively. Specifically, the HMC, ADVI meanfield and ADVI fullrank methods appear alternatively in each column of Tables (\ref{tab:propyleahmc2} and \ref{tab:propyleahmc1}) for each parameter of interest. 
Briere model has greater 95\% credible limits for $T_{min}$ while Lactin model gives negative value estimates. The results for $T_{opt}$ overlap for Bieri and Analytis in the Zero Inflated Inverse Gamma case whereas the estimates are lower for Lactin model. The $T_{max}$ credible limits overlap between Bieri and the other ecological models both in the Gaussian and the Zero Inflated Inverse Gamma case. 
The zero-rate values can be naturally modelled in the case of the Gaussian distribution. Though only the Bieri and Analytis models give wider $95\%$ credible intervals. For the rest of the models, the credible limits difference is lower than three degrees ${3^{\circ}C}$.  
In addition, in Zero Inflated Inverse Gamma case, the Analytis model has higher information criteria values, while the Bieri and Lactin models follow as shown in Table \ref{tab:perakis}, respectively. All four ecologocal models incorporate terms into their structure in the form of products which directly affect the scale of the Inverse Gamma distribution. Thus, not only the mean but also the variance of the statistical model change along and adapt the flunctuations for each temperature level as shown in Fig. \ref{fig:MFL2}. This adaptivity is evident across all the four ecological models in use. 

\begin{figure}
\begin{center}
    \includegraphics[height=6cm,width=\linewidth]{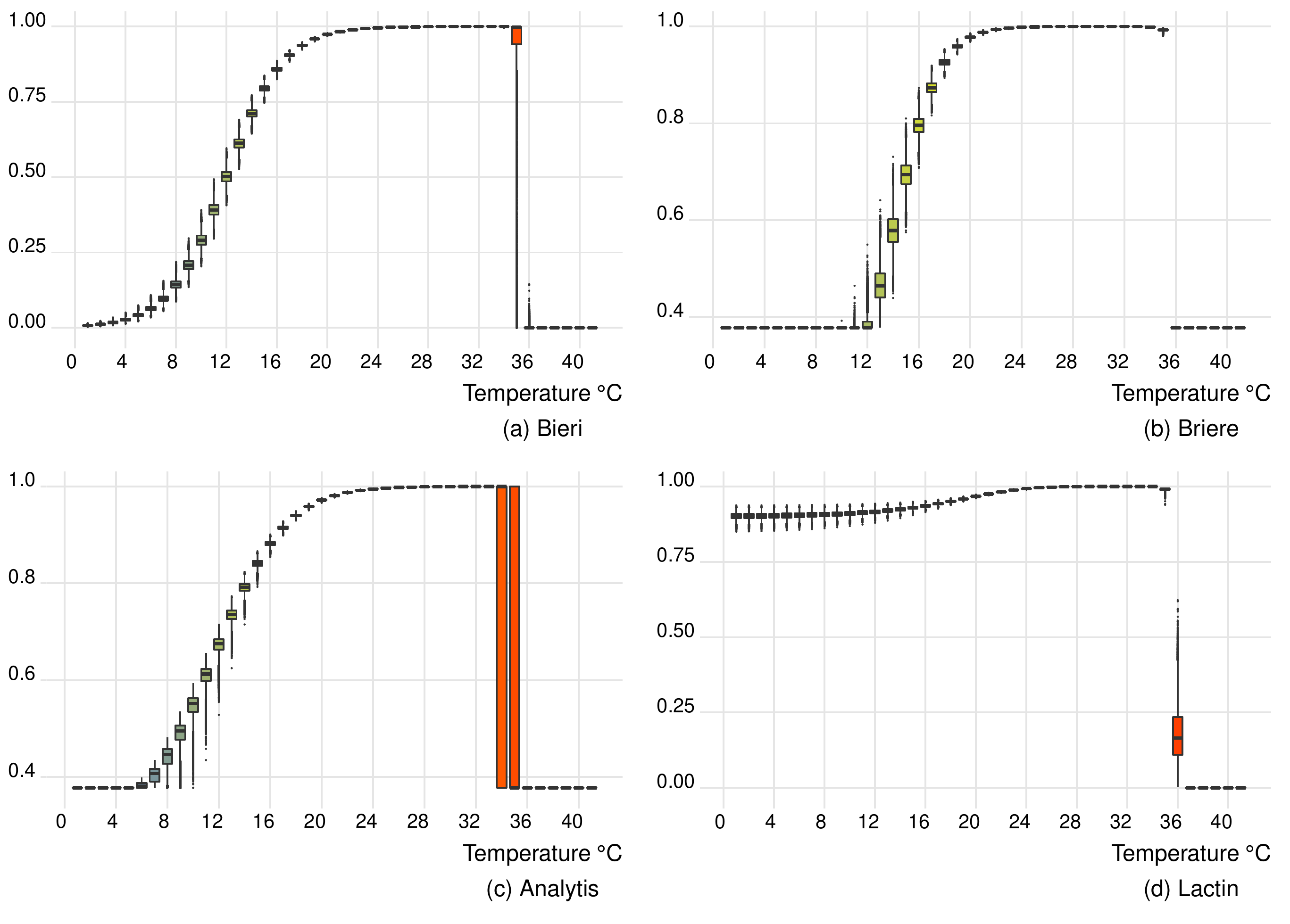}
    \caption{Boxplots of the posterior probability of non-zero development in Propylea Coccinellidae data using the Zero Inflated Inverse Gamma model}%
    \label{fig:Pzero}%
    \end{center}
\end{figure}
The $T_{min}$ outcomes for 95\% credible limits do not overlap, while Briere model has larger values. The Lactin model is not suitable to provide reasonable estimates as it allows for negative values (Table \ref{tab:propyleahmc2}).
The results for $T_{opt}$ are sorted in ascending order from  Bieri, Lactin, Analytis, and Briere. We observe that the credible limits for the $T_{max}$ overlap between models and do not exceed the maximum observed temperature $35^{o}C$.
Concerning the Zero Inflated Inverse Gamma case, the posterior predictive probabilities of non-zero entries, which are the probabilities of anthropods' development, are shown in Figure \ref{fig:Pzero}. We observe that the probabilities tend to unity near $T_{opt}$ estimates whereas become negligible out of the thermal threshold limits except for the Lactin model case which does not give robust estimates for $T_{min}$.
In addition, the 95\% credible limits estimates using ADVI meanfield and fullrank methods overlap with the HMC estimates with exceptions at the Lactin model in which the fullarank method seems to be closer to the HMC estimates (Tables \ref{tab:propyleahmc1} and \ref{tab:propyleahmc2}).
In many cases, the ADVI fullrank method agrees with HMC. It seems to give more robust estimates for $T_{max}$ mean in Gaussian case and Bieri model. However it has worst fit than HMC and also gives wider $95\%$ credible intervals.

\begin{center}
\begin{adjustbox}{{angle=0},{scale=0.6}}
\begin{threeparttable}
\caption{Posterior summaries for the four models using the Zero Inflated Inverse Gamma distribution for the Propylea Coccinellidae data. In each column we report the HMC, the ADVI-Mean field and ADVI-Full rank estimates respectively.}
\label{tab:propyleahmc2}
\begin{tabular}{|l|c|c|c|c|c|c|c|c|c|} 
\hline
                                                                                                               &          & \multicolumn{3}{c|}{$T_{min}$}& neff$^{\ddag}$                                                                                                                 &  \multicolumn{3}{c|}{$T_{opt}$}& neff$^{\ddag}$   \\ 
\hline
\multirow{4}{*}{\begin{tabular}[c]{@{}l@{}}Mean\\ \& 95\%\\Cr. I.\end{tabular}} & Bieri    & \begin{tabular}[c]{@{}c@{}}9.9 \\(9.4, 10.3)\end{tabular}          & \begin{tabular}[c]{@{}c@{}}9.4 \\(9.3, 9.6)\end{tabular}    & \begin{tabular}[c]{@{}c@{}}9.2 \\(8.7, 9.7)\end{tabular}       & 9138  & \begin{tabular}[c]{@{}c@{}}32.7 \\(32.2, 33.8)\end{tabular}       & \begin{tabular}[c]{@{}c@{}}73.0 \\32.6, 146.6\end{tabular}      & \begin{tabular}[c]{@{}c@{}}109.4 \\(28.7, 320.1)\end{tabular}     & 16552  \\ 
\cline{2-10}
                                                                                                               & Briere   & \begin{tabular}[c]{@{}c@{}}11.1 \\(10.3, 11.9)\end{tabular}        & \begin{tabular}[c]{@{}c@{}}11.2 \\(10.8, 11.5)\end{tabular} & \begin{tabular}[c]{@{}c@{}}11.2 \\(10.4, 12.0)\end{tabular}    & 14192 & \begin{tabular}[c]{@{}c@{}}29.3 \\(29.2, 29.5)\end{tabular}       & \begin{tabular}[c]{@{}c@{}}29.3 \\(29.2, 29.4)\end{tabular}       & \begin{tabular}[c]{@{}c@{}}29.3 \\(29.2, 29.5)\end{tabular}       & 17300  \\ 
\cline{2-10}
                                                                                                               & Analytis & \begin{tabular}[c]{@{}c@{}}5.0 \\(4.0, 7.0)\end{tabular}           & \begin{tabular}[c]{@{}c@{}}5.1 \\(4.1, 8.7)\end{tabular}    & \begin{tabular}[c]{@{}c@{}}4.4 \\(4.2, 4.9)\end{tabular}       & 352   & \begin{tabular}[c]{@{}c@{}}33.5 \\(32.3, 34.9)\end{tabular}       & \begin{tabular}[c]{@{}c@{}}33.0 \\(29.71, 37.0)\end{tabular}      & \begin{tabular}[c]{@{}c@{}}32.2 \\(30.5, 33.9)\end{tabular}       & 1240   \\ 
\cline{2-10}
                                                                                                               & Lactin   & \begin{tabular}[c]{@{}c@{}}-133.51 \\(-157.5, -116.8)\end{tabular} & \begin{tabular}[c]{@{}c@{}}11.6 \\(1.2, 59.2)\end{tabular}  & \begin{tabular}[c]{@{}c@{}}-12.5 \\(-409.8, 8.5)\end{tabular}  & 11511 & \begin{tabular}[c]{@{}c@{}}30.9 \\(30.8, 31.1)\end{tabular}       & \begin{tabular}[c]{@{}c@{}}39.7 \\(1.1, 211.2)\end{tabular}       & \begin{tabular}[c]{@{}c@{}}25.9 \\(5.5, 77.7)\end{tabular}        & 9847   \\ 
\hline
                                                                                                               &          & \multicolumn{3}{c|}{$T_{max}$}& neff$^{\ddag}$                                                                                                                                                 &
                          \multicolumn{3}{c|}{dev$^{\dag}$}& neff$^{\ddag}$   \\ 
\hline
\multirow{4}{*}{\begin{tabular}[c]{@{}l@{}}Mean\\ \& 95\%\\Cr. I.\end{tabular}} & Bieri    & \begin{tabular}[c]{@{}c@{}}34.4 \\(33.7, 34.9)\end{tabular}        & \begin{tabular}[c]{@{}c@{}}18.1 \\(9.3, 52.6)\end{tabular}  & \begin{tabular}[c]{@{}c@{}}16.0 \\(8.8, 52.5)\end{tabular}     & 12972 & \begin{tabular}[c]{@{}c@{}}-722.3 \\(-725.7, -714.8)\end{tabular} & \begin{tabular}[c]{@{}c@{}}-318.5 \\(-360.4, -318.8)\end{tabular} & \begin{tabular}[c]{@{}c@{}}-323.8 \\(-572.1, -309.3)\end{tabular} & 9962   \\ 
\cline{2-10}
                                                                                                               & Briere   & \begin{tabular}[c]{@{}c@{}}35.0 \\(34.8, 35.0)\end{tabular}        & \begin{tabular}[c]{@{}c@{}}34.9 \\(34.8, 35.0)\end{tabular} & \begin{tabular}[c]{@{}c@{}}34.8 \\(34.8, 35.0)\end{tabular}    & 23972 & \begin{tabular}[c]{@{}c@{}}-563.8 \\(-568.0, -555.9)\end{tabular} & \begin{tabular}[c]{@{}c@{}}-563.1 \\(-567.7, -552.0)\end{tabular} & \begin{tabular}[c]{@{}c@{}}-561.9 \\(-567.4, -549.7)\end{tabular} & 14077  \\ 
\cline{2-10}
                                                                                                               & Analytis & \begin{tabular}[c]{@{}c@{}}33.6 \\(32.5, 34.9)\end{tabular}        & \begin{tabular}[c]{@{}c@{}}33.6 \\(30.3, 37.6)\end{tabular} & \begin{tabular}[c]{@{}c@{}}32.5 \\(30.7, 34.4)\end{tabular}    & 1409  & \begin{tabular}[c]{@{}c@{}}-736.2 \\(-742.1, -729.1)\end{tabular} & \begin{tabular}[c]{@{}c@{}}-411.6 \\(-482.3, -216.9)\end{tabular} & \begin{tabular}[c]{@{}c@{}}-684.5 \\(-742.5, -618.8)\end{tabular} & 1117   \\ 
\cline{2-10}
                                                                                                               & Lactin   & \begin{tabular}[c]{@{}c@{}}34.9 \\(34.7, 35.0)\end{tabular}        & \begin{tabular}[c]{@{}c@{}}25.3 \\(1.1, 76.3)\end{tabular}  & \begin{tabular}[c]{@{}c@{}}-45.1 \\(-935.4, 52.5)\end{tabular} & 16361 & \begin{tabular}[c]{@{}c@{}}-703.6 \\(-708.1, -695.0)\end{tabular} & \begin{tabular}[c]{@{}c@{}}247.8 \\(0.0, 294.8)\end{tabular}      & \begin{tabular}[c]{@{}c@{}}247.8 \\(0.0, 294.8)\end{tabular}      & 9179   \\
\hline
\end{tabular}
\begin{tablenotes}
\item{$\dag$} deviance of the model given the data,
\item{$\ddag$} effective sample size.

\end{tablenotes}
\end{threeparttable}
\end{adjustbox}
\end{center}

Furthermore, the time elapsed until the completion of the algorithm for ADVI methods is up to 48 seconds, while for the HMC method it is at least 321 seconds for Propylea dataset as shown in Table \ref{tab:etime1}. The average difference between the working times of HMS and each ADVI method is around 6300 seconds (105 minutes) whereas between the meanfield and fullrank is around 12 seconds. These findings illuminate both the VBI fullrank method's computational time-efficacy. 

Bayesian model averaging results are shown in Tables  \ref{tab:propyleaw}, \ref{tab:bma}. This time the predictive bias that affect the model averaging estimates are from the Bieri and Analytis models fot he Gaussian and the Zero Inflated Inverse Gamma distribution respectively.

\section{Conclusions}
\label{results}
Conclusions of the current work can be summarised into five groups.

\subsection{Comparison of Computational Methods}
Although ADVI methods agree with HMC in many of cases and are distinguished in time-efficacy, robust estimates do not appear to be available for all parameters of interest, with the ADVI fullrank method providing better estimates than the ADVI meanfield mehod. The two datasets application show that they are sensitive to the initialization of the parameters and the root-finding algorithm used, in particular for more complicated models such as Analytis and Lactin. For example, if the root initial suggestion is positive, the root generated is also positive, even if the HMC results are negative, like in Tables(\ref{tab:propyleahmc4}, \ref{tab:propyleahmc2}, \ref{tab:propyleahmc1}). However, in the same tables, the VBI fullrank method's credible limits overlap with those of the HMC method more frequently than the VBI meanfield method, indicating that the ADVI fullrank method has little better computational efficiency.

\subsection{Distribution of the data}
Inverse Gamma distribution is a promising alternative to the classical Gaussian distribution in the analysis of the rate of development of anthropods in the absence of zero data and also provides adaptive temperature-level variance predictions across all ecological models. It also reinforces the flexibility of the Lactin model. In addition Inverse Gamma distribution naturally models the developmental rates that are defined as the reciprocal of positive real numbers.  

\subsection{Zero-inflation performance}
Zero Inflated Inverse Gamma distribution can be used in the presence of non-development (zero data) after defining the probability of generating zeros. In this way the developmental rates are naturally modeled and also the predictions are variance adaptive to the temperature levels. Finally, it performs better than the classical Gaussian model in the Analytis case as far as information criteria scores and marginal likelihood estimates are concerned. 

\subsection{Model comparison}
For the model comparison, both the information criteria and the marginal probability values agree on the prioritization of performance across ecological models at both data-sets. Nevertheless, there is some variation between marginal likelihood estimates within ecological models and their is no clear pattern across the marginal likelihood approaches.

There is no standard ecological model that has better fit measured by information criteria and the estimated marginal likelihood than the others in all the situations. However, even in the presence of zero response values, the Bieri ecological model appears to have robust results using the Gaussian distribution and stands out when there is a linear increase of the developmental rate like in the \textit{Propylea quatuordecimpunctata} data-set case. In addition, it provides interpretable estimates of interest parameters with credible intervals, often overlapping with that of the Analytis or the Lactin model, using weakly informative priors. By contrast, the Bieri model can not follow the non-linear increase in the rate of development of the \textit{Tetranychus} data, as do the more sophisticated Analytis or / and Lactin models.\\ Lactin model has better fit in the case of exponentially fluctuations of the developmental rate, but is inadequate for $T_{min}$ estimation. Analytis model needs tuning for $T_{min}$ sample space so as not to underestimate it and provides robust estimates for the other thermal parameters involved in the model.

Briere and Analytis models by definition includes the true threshold parameters $T_{min}$ and $T_{max}$ along with a functional truncation between these threshold parameters, which do not affect their performance. Moreover, the Briere model appears to have the worst marginal likelihood values, as well as the worst predictive performance for both Gaussian and Inverse Gamma distributions, despite having the least parameters and being widely used. It also seems to generate estimates with small 95\% credible intervals and higher $T_{max}$ values. Although statistical inference based on the information criteria have reasonable values for Briere model, when we focus on a particular parameter of interest such as the $T_{max}$, its performance may be poor compared to the other ecological models as shown in Fig. \ref{fig:MFL4}. In conclusion, the Briere model can be used to obtain quick estimates of thermal parameters involved in arthropod evolution without the need for any pre-adjustment (parameters tuning or / and transformation).

The Analytis model needs some initial boost by either tuning the thermal parameters or using informative priors in order to achieve good predictions. In particular, a lower bound of $4^{o}C$ for $T_{min}$ should be imposed in order not to generate negligible or negative values. An upper bound for $T_{max}$ should also be imposed if there are zero responses at a higher temperature. It is only after these parameters have been calibrated that the Analytis model stands out in its performance by comparing the other models and generates wide 95\% credible intervals for parameters of interest. Additionally, due to its functional form, Analytis model can fit any polynomial shaped increase and decrease of the developmental rate. In comparison to the simple Gaussian case, Analytis explains the zeros generated scheme better. However, it provides smaller estimates than the other three models, for the $T_{min}$ parameter. In summary, the Analytis model performs well to predict the upper thermal parameters only after adjusting the $T_{min}$ and $T_{max}$ sample space interval limits to prior information from the data.

In addition, the Lactin model does not include an analytical lower thermal threshold in its form. The lower thermal threshold estimates are implicitly derived as the lower root of the Lactin model applying the appropriate model parameters values. Thus, the Lactin model in some cases generates negative estimates of $T_{min}$ that can not be interpreted in ecology studies. Also parameters $\lambda$ and $\varDelta$ fluctuations trigger statistical model discontinuity situations that need to be addressed taking into account the data-set structure. Moreover, it has smaller effective sample size especially in the case of Gaussian distribution. On the contrary, it generates a wider 95\% credible interval for $T_{max}$ compared to other ecological models and also performs better in the case of zero responses at high temperatures. In summary, the Lactin model is applicable when research focus on robust estimates of the optimum and the upper thermal threshold parameters without the need for prior tuning. 

\subsection{BMA performance}
BMA technique is used to  address the difficulty that the true causal structure in ecological data is often not known and often several models are used to describe the development of anthropods. The mean square error in model-averaged predictions depends on each model’s  predictive bias and variance, the covariance in predictions between models, and the uncertainty about model weights \citep{dormann2018model}. In this work, we adopt weights in (\ref{IC_weights}). The BMA weights summarised in Tables(\ref{tab:tetraw}, \ref{tab:propyleaw}) clearly support the Bieri and Analytis models for the Propylea data and the Analytis and Lactin models for the Tetranychus data. As a result, the BMA estimates are close to the estimates of these models respectively for the two data-sets. 

\section{Discussion}

There are several issues that we deal with in terms of arthropod developmental rates, and there are some concerns for future research. To begin with, comparing non-linear non-nested models with varying numbers of parameters and truncated mean structures, as well as excessive-zeros in data, is not a trivial task.
\subsection{Comparison of Computational Methods}
We use the Bayesian paradigm, as well as some contemporary computational approaches such as the HMC, ADVI-meanfield, and ADVI-fullrank, to address not only irregular and truncated mean structures, but also the uncertainty of zero generation in the data. Although ADVI techniques are gaining popularity in the scientific community due to their fast and computationally inexpensive approximations to the posterior distributions, they do not provide robust estimates of all the parameters of the models we study. The HMC method, on the other hand, gives robust estimates even under these specific model and data structure conditions.
\subsection{Distribution of the data}
Furthermore, for the data generation scheme, we suggest the Gaussian and the Inverse Gamma distributions. The Gaussian option provides sensible estimates that can be used even though the data has a lot of zeros. Inverse Gamma, on the other hand, not only naturally models developmental rates, which are characterized as the reciprocal of positive real values, but also provides variance adaptivity across temperature fluctuations. Allso for each ecological model we define the Zero Inflated Inverse Gamma density so as to model data with an excessive number of zeros. When comparing models involving Gaussian and Inverse Gamma or Zero Inflated gamma distributions, we find that the second performs better in non-zero data cases, while the first performs better in all but the Analytis model.
\subsection{Model comparison}
In addition, we address the model comparison challenge by employing the Information criteria not only to assess model goodness of fit to data while accounting for model complexity, but also to assess ability of models to make predictions on new data using the leave one out cross validation technique. Additionally, we use marginal likelihood approximations of the various models to determine which one is best supported by the data. Finally, we plot the predicted posteriors alongside the observed data points to visualize the prediction ability of the suggested models.
\subsection{BMA performance}
The predictive bias of the weighted models, as well as the uncertainty about the weights, affect BMA results using Information criteria as weights according to (\ref{IC_weights}) weights as outlined in section \ref{BMA_theory} in the appendix. As a result, the BMA approach does not provide estimates that differ from the best-performing models.
\subsection{Future research}
Among things for future research is to select consistently the most robust candidate between models given sufficiently many data samples, in a sensitivity analysis perspective. Moreover, the ADVI methods, can be extended so as to capture more sophisticated mean structures, like the ones we present in the current work. In addition, probability density that generates zeros in the Zero Inflated Inverse Gamma distribution can be modeled in more complex ways, such as using hyperparameters and hierarchical effects across temperature levels. Finally, R-packages that include the suggested models and perform the analysis presented in this paper are to be created.

\bibliographystyle{rss}
\bibliography{main}
\clearpage

\setcounter{table}{0}
\renewcommand{\thetable}{A\arabic{table}}
\appendix
\renewcommand*\appendixpagename{\Large Appendices}
\appendixpage
\section{Tables with summaries and posterior estimates of the statistical methods used}
\subsection{\textit{Propylea quatuordecimpunctata} dataset}

\begin{center}
\begin{threeparttable}
\centering
\captionsetup{width=1\textwidth}
\caption{Algorithmic working time in seconds, for two datasets.}
\label{tab:etime1}
\sisetup{table-number-alignment = center,round-mode = figures,
round-precision = 3}
  \begin{tabular}{|c|c|c|c|c|}
  \multicolumn{5}{c}{{\textit{Tetranychus urticae}}}\\\hline
  \multicolumn{5}{c}{{Gaussian}}\\ \hline

  \hline
        & Bieri  & Briere & Analytis & Lactin \\
  \hline
HMC          & 2011      & 476                       & 9852
 & 5672  \\
ADVI-meanfield          & 8 & 1 & 52 & 1 \\
ADVI-fullrank        & 45 & 6 & 38 & 44 \\
\hline
\multicolumn{5}{c}{{Inverse Gamma}}\\ \hline

HMC          & 1375 & 761 & 3231 & 4917 \\
ADVI-meanfield          & 2 & 4 & 15 & 27 \\
ADVI-fullrank        & 2 & 34 & 14 & 44 \\ \hline
  \multicolumn{5}{c}{{\textit{ }}}\\[0.01cm]
  \multicolumn{5}{c}{{\textit{Propylea quatuordecimpunctata}}}\\ \hline
  \multicolumn{5}{c}{{Gaussian}}\\ \hline
  \hline
        & Bieri  & Briere & Analytis & Lactin \\
  \hline
HMC          & 4335      & 20218                       & 4913
 & 1426  \\
ADVI-meanfield          & 13                       & 7 & 3 & 16 \\
ADVI-fullrank        & 48                       & 11 & 12 & 16 \\
\hline
\multicolumn{5}{c}{{Zero Inflated Inverse Gamma}}\\ \hline

HMC          & 1078                       & 321 & 15291 & 3073 \\
ADVI-meanfield          & 41                       & 5 & 9 & 10 \\
ADVI-fullrank        & 52                       & 5 & 39 & 18 \\ \hline
  
\end{tabular}
\end{threeparttable}
\end{center}

\newpage

\begin{center}
\begin{threeparttable}
\caption{BMA weights for the \textit{Tetranychus urticae} data}
\label{tab:tetraw}
\sisetup{table-number-alignment = center,round-mode = figures,
round-precision = 3}
\begin{tabular}{|c|c|c|c|c|c|c|c|}
\hline
\multicolumn{8}{|c|}{Gaussian distribution}                                       \\ \hline
         & aic\_w  & dic\_w  & loocv\_w & waic\_w & bic\_w  & elbo\_mf & elbo\_fr \\ \hline
Bieri   & 4.9E-17 & 9.4E-16 & 2.3E-16  & 2.4E-16 & 8.1E-18 & 3.1E-53 & 7.3E-45 \\ \hline
Briere   & 1.9E-32 & 2.3E-31 & 6.6E-33  & 6.9E-33 & 1.1E-31 & 2.9E-27  & 1.5E-18  \\ \hline
Analytis & 1 & 1 & 1  & 1 & 1 & 0.001  & 1  \\ \hline
Lactin   & 1.0E-14 & 4.9E-14 & 1.2E-12  & 1.2E-12 & 9.0E-18 & 0.999  & 3.5E-39    \\ 
\hline
\multicolumn{8}{|c|}{Inverse Gamma distribution}                                  \\ \hline
Bieri   & 3.0E-43 & 4.5E-43 & 1.1E-42  & 1.0E-42 & 3.0E-43 & 1.6E-20 & 1.0E-200\\ \hline
Briere   & 6.6E-36 & 8.1E-36 & 2.2E-36  & 2.1E-36 & 3.8E-35 & 8.9E-19  & 1.0E-171  \\ \hline
Analytis & 1.4E-5 & 7.1E-6 & 2.1E-5  & 2.1E-5 & 1.9E-6 & 1  & 1  \\ \hline
Lactin   & 0.999 & 0.999 & 0.999  & 0.999 & 1 & 1.0E-36  & 2.7E-58    \\ \hline
\end{tabular}
\end{threeparttable}
\end{center}

\begin{center}
\begin{threeparttable}
\caption{BMA weights for the \textit{Propylea quatuordecimpunctata} data}
\label{tab:propyleaw}
\sisetup{table-number-alignment = center,round-mode = figures,
round-precision = 3}
\begin{tabular}{|c|c|c|c|c|c|c|c|}
\hline
\multicolumn{8}{|c|}{Gaussian distribution}                                        \\ \hline
         & aic\_w  & dic\_w   & loocv\_w & waic\_w & bic\_w  & elbo\_mf & elbo\_fr \\ \hline
Bieri   & 0.999 & 0.968 & 0.991  & 0.991 & 1 & 1  & 1  \\
Briere   & 3.3E-34 & 7.3E-30 & 5.5E-35  & 5.5E-35 & 4.8E-33 & 5E-203  & 3E-218  \\
Analytis & 1.1E-31 & 1.5E-32 & 3.8E-32  & 3.8E-32 & 4.0E-31 & 9.0E-269  & 1.0E-210  \\
Lactin   & 5.9E-04 & 0.032 & 0.009  & 0.009 & 1.1E-05 & 5.0E-159  & 3.0E-162  \\ \hline
\multicolumn{8}{|c|}{Zero Inflated Inverse Gamma distribution}                                   \\ \hline
Bieri   & 9.3E-5 & 0.001 & 0.001  & 0.001 & 3.5E-4 & 5.1E-53 & 3.4E-72 \\
Briere   & 1.7E-38 & 3.9E-38 & 7.3E-38  & 7.2E-38 & 2.4E-37 & 1  & 1.1E-23  \\
Analytis & 1 & 0.999 & 0.999  & 0.999 & 1 & 3.3E-35  & 1  \\
Lactin   & 1.7E-8 & 2.1E-8 & 5.8E-8  & 5.8E-8 & 6.4E-8 & 3.0E-114  & 2.0E-140  \\ \hline
\end{tabular}
\end{threeparttable}
\end{center}

\newpage
\begin{adjustbox}{{angle=0},{scale=0.6}}
\begin{threeparttable}
\caption{The mean and $95\%$ Cr.I. limits of the BMA estimates of parameters of interest calculated with the use of information criteria score weights assuming the Gauussian and Inverse Gamma distributions respectively for \textit{Tetranychus urticae} data.}
\label{tab:bma_tetra}
\sisetup{table-number-alignment = center,round-mode = figures,
round-precision = 3}
\begin{tabular}{|c|c|c|c|c|c|c|c|c|}
\hline
\multicolumn{9}{|c|}{Gaussian Model}                                                                                                                                                 \\ \hline
          &  & aic\_w                & dic\_w                & loocv\_w              & waic\_w               & bic\_w                & elbo\_mf              & elbo\_fr              \\ \hline
$T_{min}$ & Mean  & 4.4                 & 4.4                 & 4.4                 & 4.4                 & 4.4                 & 4.4                 & 4.4                 \\ \hline
          & $95\%$ Cr.I. & (4.0, 5.3)         & (4.0, 5.3)         & (4.0, 5.3)         & (4.0, 5.3)         & (4.0, 5.3)         & (4.0, 5.3(         & (4.0, 5.3)         \\ \hline
$T_{opt}$    &  Mean & 32.9                & 32.9                & 32.9                & 32.9                & 32.9                & 32.6                & 32.9                \\ \hline
          & $95\%$ Cr.I. & (32.7, 33.3)       & (32.7, 33.3)       & (32.7, 33.3)       & (32.7, 33.3)       & (32.7, 33.3)       & (32.4, 32.8)        & (32.7, 33.3)        \\ \hline
$T_{max}$    & Mean & 35.3                & 35.3                & 35.3                & 35.3                & 35.3                & 36.9                & 35.3                \\ \hline
          & $95\%$ Cr.I. & (35.1, 35.5)         & (35.1, 35.5)         & (35.1, 35.5)         & (35.1, 35.5)         & (35.1, 35.5)       & (36.5, 37.3)       & (35.1, 35.5)       \\ \hline
dev$^{\dag}$       & Mean & -1750.3             & -1750.3             & -1750.3             & -1750.3             & -1750.3             & -1693.9             & -1750.3             \\
        \hline
          & $95\%$ Cr.I. & (-1758.0,-1738.3) & (-1758.0,-1738.3) & (-1758.0,-1738.3) & (-1758.0,-1738.3) & (-1758.0,-1738.3) & (-1703.1,-1679.7) & (-1758.0,-1738.3) \\ \hline
\multicolumn{9}{|c|}{Inverse Gamma Model}                                              \\ \hline
$T_{min}$ & Mean & 4.2                 & 4.2                 & 4.2                 & 4.2                 & 4.2                 & 4.2                 & 4.2                 \\
        \hline
         & $95\%$ Cr.I.  & (4.0, 4.6)         & (4.0, 4.6)         & (4.0, 4.6)         & (4.0, 4.6)         & (4.0, 4.6)         & (4.0, 4.6)         & (4.0, 4.6)         \\ \hline
$T_{opt}$    & Mean T\_opt & 32.0                & 32.0                & 32.0                & 32.0                & 32.0                & 33.6                & 33.6                \\
        \hline
          & $95\%$ Cr.I. & (31.8, 32.2)       & (31.8, 32.2)       & (31.8, 32.2)       & (31.8, 32.2)       & (31.8, 32.2)       & (33.3, 33.9)       & (33.3, 33.9)       \\ \hline
$T_{max}$    & Mean & 38.4                & 38.4                & 38.4                & 38.4                & 38.4                & 35.0                & 35.0                \\        \hline
          & $95\%$ Cr.I. & (38.1, 38.9)       & (38.1, 38.9)       & (38.1, 38.9)       & (38.1, 38.9)       & (38.1, 38.9)       & (35.0, 35.1)       & (35.0, 35.1)       \\ \hline
dev$^{\dag}$       & Mean & -1916.3 & -1916.3 & -1916.3 & -1916.3 & -1916.3 & -1894.0             & -1894.0             \\ \hline
          & $95\%$ Cr.I. & (-1920.6,-1908.3) & (-1920.6,-1908.3) & (-1920.6,-1908.3) & (-1920.6,-1908.3) & (-1920.6,-1908.3) & (-1899.2,-1885.2) & (-1899.2,-1885.2) \\ \hline
\end{tabular}
\begin{tablenotes}
\item{$\dag$} deviance of the model given the data.
\end{tablenotes}
\end{threeparttable}
\end{adjustbox}

\begin{center}
\begin{adjustbox}{{angle=0},{scale=0.65}}
\begin{threeparttable}
\caption{The mean and $95\%$ Cr.I. limits of the BMA estimates of parameters of interest calculated with the use of information criteria score weights assuming the Gaussian and Inverse Gamma distributions respectively for the \textit{Tetranychus urticae} data.}
\label{tab:bma}
\sisetup{table-number-alignment = center,round-mode = figures,
round-precision = 3}
\begin{tabular}{|c|c|c|c|c|c|c|c|c|}
\hline
\multicolumn{9}{|c|}{Gaussian distribution}                                                                                                                                   \\ \hline
       &            & aic\_w              & dic\_w              & loocv\_w            & waic\_w             & bic\_w              & elbo\_mf            & elbo\_fr            \\ \hline
$T_{min}$ & Mean       & 10.6              & 10.6              & 10.6              & 10.6              & 10.6              & 10.6              & 10.6     \\  \hline
       & $95\%$ Cr.I. & (9.7, 11.4) & (9.7, 11.4) & (9.7, 11.4) & (9.7, 11.4)      & (9.7, 11.4) & (9.7, 11.4) & (9.7, 11.4)            \\ \hline
$T_{opt}$ & Mean       & 32.6              & 32.5              & 32.6              & 32.6              & 32.6              & 32.6              & 32.6              \\ \hline
       & $95\%$ Cr.I. & (32.0, 33.5)     & (31.9, 33.4)     & (32.0, 33.5)      & (32.0, 33.5)      & (32.0, 33.5)     & (32.0, 33.5)     & (32.0, 33.5)         \\ \hline
$T_{max}$ & Mean      & 35.0              & 35.0              & 35.0              & 35.0              & 35.0              & 35.0              & 35.0        \\ \hline
       & $95\%$ Cr.I. & (34.98, 35.0)     & (34.98, 35.0)     & (34.98, 35.0)     & (34.98, 35.0)     & (34.98, 35.0)     & (34.98, 35.0)     & (34.98, 35.0)     \\ \hline
dev$^{\dag}$    & Mean       & -818.8            & -818.5            & -818.7            & -818.7            & -818.8            & -818.8            & -818.8           \\ \hline
       & $95\%$ Cr.I. & (-833.1, -800.8) & (-832.4, -801.1) & (-832.9, -800.9)  & (-832.9, -800.9)  & (-833.1, -800.8) & (-833.1, -800.8) & (-833.1, -800.8)          \\ \hline
\multicolumn{9}{|c|}{Inverse Gamma distribution}                                                                                                                              \\ \hline
$T_{min}$ & Mean       & 5.0               & 5.0               & 5.0              & 5.0              & 5.0               & 11.1               & 5.0          \\ \hline
       & $95\%$ Cr.I.  & (4.0, 7.0)        & (4.0, 7.0)        & (4.0, 7.0)      & (4.0, 7.0)      & (4.0, 7.0)        & (10.3, 11.9)        & (4.0, 7.0)          \\ \hline
$T_{opt}$ & Mean       & 33.5              & 33.5              & 33.5              & 33.5              & 33.5              & 29.3              & 33.5              \\ \hline
       & $95\%$ Cr.I. & (32.3, 34.9) & (32.3, 34.9) & (32.3, 34.9) & (32.3, 34.9) & (32.3, 34.9) & (29.2, 29.5) & (32.3, 34.9)         \\ \hline
$T_{max}$ & Mean       & 33.6              & 33.6              & 33.6              & 33.6              & 33.6              & 35.0              & 33.6          \\ \hline
       & $95\%$ Cr.I. & (32.5, 34.9)     & (32.5, 34.9)     & (32.5, 34.9)     & (32.5, 34.9)     & (32.5, 34.9)     & (34.8, 35.0)     & (32.5, 34.9)           \\ \hline
dev$^{\dag}$    & Mean       & -736.2            & -736.2            & -736.2            & -736.2            & -736.2            & -563.8            & -736.2                 \\ \hline
       & $95\%$ Cr.I. & (-742.1, -729.1) & (-742.1, -729.1) & (-742.0, -729.1) & (-742.0, -729.1) & (-742.1, -729.1) & (-568.0, -555.9) & (-742.1, -729.1)   \\ \hline
\end{tabular}
\begin{tablenotes}
\item{$\dag$} deviance of the model given the data.
\end{tablenotes}
\end{threeparttable}
\end{adjustbox}
\end{center}

\begin{center}
\begin{adjustbox}{{angle=0},{scale=0.60}}
\begin{threeparttable}
\caption{Posterior summaries for the four models using the Gaussian distribution for the Propylea Coccinellidae data. In each column we report the HMC, the ADVI-Mean field and ADVI-Full rank estimates respectively.}
\label{tab:propyleahmc1}
\begin{tabular}{|c|c|c|c|c|c|c|c|c|c|} 
\hline
                                                                              &          & \multicolumn{3}{c|}{$T_{min}$}& neff$^{\ddag}$                                                                                                                 &  \multicolumn{3}{c|}{$T_{opt}$}& neff$^{\ddag}$         \\ 
\hline
\multirow{4}{*}{\begin{tabular}[c]{@{}c@{}}Mean\\ 95\%\\ Cr. I.\end{tabular}} & Bieri    & \begin{tabular}[c]{@{}c@{}}10.6 \\(9.7, 11.4)\end{tabular}        & \begin{tabular}[c]{@{}c@{}}10.7 \\(10.6, 10.8)\end{tabular}       & \begin{tabular}[c]{@{}c@{}}10.7 \\(10, 11.4)\end{tabular}         & 7906                      & \begin{tabular}[c]{@{}c@{}}32.6 \\(32, 33.5)\end{tabular}         & \begin{tabular}[c]{@{}c@{}}32.2 \\(32, 32.5)\end{tabular}         & \begin{tabular}[c]{@{}c@{}}32.2 \\(31.8, 32.5)\end{tabular}       & 7620                       \\ 
\cline{2-10}
                                                                              & Briere   & \begin{tabular}[c]{@{}c@{}}13.1 \\(12.4, 13.8)\end{tabular}       & \begin{tabular}[c]{@{}c@{}}13.3 \\(12.9, 13.8)\end{tabular}       & \begin{tabular}[c]{@{}c@{}}8.2 \\(1.8, 15.6)\end{tabular}         & 4                         & \begin{tabular}[c]{@{}c@{}}29.7 \\(29.6, 29.8)\end{tabular}       & \begin{tabular}[c]{@{}c@{}}29.7 \\(29.5, 29.8)\end{tabular}       & \begin{tabular}[c]{@{}c@{}}27.3 \\(11.7, 30.1)\end{tabular}       & 4                          \\ 
\cline{2-10}
                                                                              & Analytis & \begin{tabular}[c]{@{}c@{}}6.3 \\(4.1, 10.6)\end{tabular}         & \begin{tabular}[c]{@{}c@{}}21.5 \\(7.7, 55.2)\end{tabular}        & \begin{tabular}[c]{@{}c@{}}6.9 \\(5.4, 9.3)\end{tabular}          & 3742                      & \begin{tabular}[c]{@{}c@{}}33.2 \\(32.1, 34.8)\end{tabular}       & \begin{tabular}[c]{@{}c@{}}39.2 \\(18.1, 89.1)\end{tabular}       & \begin{tabular}[c]{@{}c@{}}33.2 \\(31.7, 37.7)\end{tabular}       & 4435                       \\ 
\cline{2-10}
                                                                              & Lactin   & \begin{tabular}[c]{@{}c@{}}-154.4 \\(-234.7, -119.3)\end{tabular} & \begin{tabular}[c]{@{}c@{}}-146.2 \\(-154.3, -138.1)\end{tabular} & \begin{tabular}[c]{@{}c@{}}-159.2 \\(-243.1, -119.0)\end{tabular} & 7560                      & \begin{tabular}[c]{@{}c@{}}31.2 \\(30.9, 31.3)\end{tabular}       & \begin{tabular}[c]{@{}c@{}}31.1 \\(31.1, 31.2)\end{tabular}       & \begin{tabular}[c]{@{}c@{}}31.1 \\(30.9, 31.3)\end{tabular}       & 8024                       \\ 
\hline
                                                                              &          & \multicolumn{3}{c|}{$T_{max}$}& neff$^{\ddag}$                                                                                                                                                 &
                          \multicolumn{3}{c|}{dev$^{\dag}$}& neff$^{\ddag}$ \\ 
\hline
\multirow{4}{*}{\begin{tabular}[c]{@{}c@{}}Mean\\ 95\%\\ Cr. I.\end{tabular}} & Bieri   & \begin{tabular}[c]{@{}c@{}}35.0 \\(34.98, 35.02)\end{tabular}     & \begin{tabular}[c]{@{}c@{}}33.8 \\(33.6, 33.9)\end{tabular}       & \begin{tabular}[c]{@{}c@{}}33.7 \\(33.5, 33.9)\end{tabular}       & 7194                      & \begin{tabular}[c]{@{}c@{}}-818.7 \\(-833.1, -800.6)\end{tabular} & \begin{tabular}[c]{@{}c@{}}-672.7 \\(-685.1, -657.4)\end{tabular} & \begin{tabular}[c]{@{}c@{}}-672.5 \\(-686.2, -657.1)\end{tabular} & 11673                      \\ 
\cline{2-10}
                                                                              & Briere   & \begin{tabular}[c]{@{}c@{}}35.0 \\(35.0, 35.01)\end{tabular}      & \begin{tabular}[c]{@{}c@{}}35.0 \\(34.8, 35)\end{tabular}         & \begin{tabular}[c]{@{}c@{}}32.9 \\(14.3, 35)\end{tabular}         & 113                       & \begin{tabular}[c]{@{}c@{}}-660.5 \\(-667.0, -649.7)\end{tabular} & \begin{tabular}[c]{@{}c@{}}-535.8 \\(-543.2, -520.5)\end{tabular} & \begin{tabular}[c]{@{}c@{}}-429.7 \\(-535.0, -402.1)\end{tabular} & 14478                      \\ 
\cline{2-10}
                                                                              & Analytis & \begin{tabular}[c]{@{}c@{}}33.6 \\(32.5, 34.9)\end{tabular}       & \begin{tabular}[c]{@{}c@{}}42.5 \\(32.7, 94.5)\end{tabular}       & \begin{tabular}[c]{@{}c@{}}36.1 \\(4.6, 133.2)\end{tabular}       & 9370                      & \begin{tabular}[c]{@{}c@{}}-674.1 \\(-690.2, -666.3)\end{tabular} & \begin{tabular}[c]{@{}c@{}}-160.0 \\(-261.3, -98.2)\end{tabular}  & \begin{tabular}[c]{@{}c@{}}-491.9 \\(-595.9, -293.6)\end{tabular} & 5164                       \\ 
\cline{2-10}
                                                                              & Lactin   & \begin{tabular}[c]{@{}c@{}}35.0 \\(35, 35.04)\end{tabular}        & \begin{tabular}[c]{@{}c@{}}35.0 \\(35.0, 35.1)\end{tabular}       & \begin{tabular}[c]{@{}c@{}}35.0 \\(35.0, 35.1)\end{tabular}       & 21622                     & \begin{tabular}[c]{@{}c@{}}-809.8 \\(-823.7, -792.2)\end{tabular} & \begin{tabular}[c]{@{}c@{}}-804.7 \\(-820.5, -784.8)\end{tabular} & \begin{tabular}[c]{@{}c@{}}-778.5 \\(-807.5, -702.6)\end{tabular} & 12147                      \\
\hline
\end{tabular}
\begin{tablenotes}
\item{$\dag$} deviance of the model given the data,
\item{$\ddag$} effective sample size.
\end{tablenotes}
\end{threeparttable}
\end{adjustbox}
\end{center}

\begin{adjustbox}{{angle=0},{scale=0.65}}
\begin{threeparttable}
\caption{Posterior summaries for the four models using the Gaussian distribution for the \textit{Tetranychus urticae} data. In each column we report the HMC, the ADVI-Mean field and ADVI-Full rank estimates respectively.}
\label{tab:propyleahmc3}
\begin{tabular}{|c|c|c|c|c|c|c|c|c|} 
\hline
         & \multicolumn{3}{c|}{$T_{min}$}& neff$^{\ddag}$                                                                                                                 &  \multicolumn{3}{c|}{$T_{opt}$}& neff$^{\ddag}$   \\ 
\hline
Bieri    & \begin{tabular}[c]{@{}c@{}}10.5 \\(10.2, 10.8)\end{tabular} & \begin{tabular}[c]{@{}c@{}}9.8 \\(9.5, 10.2)\end{tabular}   & \begin{tabular}[c]{@{}c@{}}9.8 \\(9.2, 10.3)\end{tabular}   & 15089                     & \begin{tabular}[c]{@{}c@{}}33.0 \\(32.7, 33.4)\end{tabular}          & \begin{tabular}[c]{@{}c@{}}158.6 \\(144.4, 165.4)\end{tabular}       & \begin{tabular}[c]{@{}c@{}}183.8 \\(76.4, 368.5)\end{tabular}        & 8164                       \\ 
\hline
Briere   & \begin{tabular}[c]{@{}c@{}}9.3 \\(8.6, 9.9)\end{tabular}    & \begin{tabular}[c]{@{}c@{}}9.3 \\(9.1, 9.5)\end{tabular}    & \begin{tabular}[c]{@{}c@{}}9.3 \\(8.6, 9.9)\end{tabular}    & 8690                      & \begin{tabular}[c]{@{}c@{}}33.1 \\(32.5, 33.7)\end{tabular}          & \begin{tabular}[c]{@{}c@{}}33.0 \\(32.8, 33.2)\end{tabular}          & \begin{tabular}[c]{@{}c@{}}33.1 \\(32.4, 33.7)\end{tabular}          & 8028                       \\ 
\hline
Analytis & \begin{tabular}[c]{@{}c@{}}4.4 \\(4.0, 5.3)\end{tabular}    & \begin{tabular}[c]{@{}c@{}}4.3 \\(4.2, 4.4)\end{tabular}    & \begin{tabular}[c]{@{}c@{}}5.3 \\(4.8, 6.1)\end{tabular}    & 8757                      & \begin{tabular}[c]{@{}c@{}}32.9 \\(32.7, 33.3)\end{tabular}          & \begin{tabular}[c]{@{}c@{}}33.0 \\(32.9, 33.1)\end{tabular}          & \begin{tabular}[c]{@{}c@{}}33.2 \\(32.6, 33.7)\end{tabular}          & 41                         \\ 
\hline
Lactin   & \begin{tabular}[c]{@{}c@{}}10.4 \\(10.1, 10.7)\end{tabular} & \begin{tabular}[c]{@{}c@{}}-3.9 \\(-5.8, -1.9)\end{tabular} & \begin{tabular}[c]{@{}c@{}}-0.7 \\(-4.0, 3.0)\end{tabular}  & 11526                     & \begin{tabular}[c]{@{}c@{}}32.6 \\(32.4, 32.8)\end{tabular}          & \begin{tabular}[c]{@{}c@{}}32.2 \\(32.0, 32.3)\end{tabular}          & \begin{tabular}[c]{@{}c@{}}32.0 \\(31.8, 32.2)\end{tabular}          & 9884                       \\ 
\hline
         & \multicolumn{3}{c|}{$T_{max}$}& neff$^{\ddag}$                                                                                                                                                 &
                          \multicolumn{3}{c|}{dev$^{\dag}$}& neff$^{\ddag}$  \\ 
\hline
Bieri    & \begin{tabular}[c]{@{}c@{}}36.3 \\(35.9, 36.8)\end{tabular} & \begin{tabular}[c]{@{}c@{}}164.9 \\(161, 169)\end{tabular}  & \begin{tabular}[c]{@{}c@{}}190.6 \\(80, 376.1)\end{tabular} & 7561                      & \begin{tabular}[c]{@{}c@{}}-1677.1 \\(-1683.2, -1666.6)\end{tabular} & \begin{tabular}[c]{@{}c@{}}-1461.0 \\(-1467.7, -1443.6)\end{tabular} & \begin{tabular}[c]{@{}c@{}}-1461.7 \\(-1467.7, -1456.3)\end{tabular} & 10087                      \\ 
\hline
Briere   & \begin{tabular}[c]{@{}c@{}}40.0 \\(39.3, 40.9)\end{tabular} & \begin{tabular}[c]{@{}c@{}}39.9 \\(39.6, 40.1)\end{tabular} & \begin{tabular}[c]{@{}c@{}}40.0 \\(39.1, 40.9)\end{tabular} & 7562                      & \begin{tabular}[c]{@{}c@{}}-1602.2 \\(-1606.8, -1593.4)\end{tabular} & \begin{tabular}[c]{@{}c@{}}-1593.2 \\(-1605.7, -1569.1)\end{tabular} & \begin{tabular}[c]{@{}c@{}}-1598.5 \\(-1606.3, -1583.6)\end{tabular} & 9618                       \\ 
\hline
Analytis & \begin{tabular}[c]{@{}c@{}}35.3 \\(35.1, 35.5)\end{tabular} & \begin{tabular}[c]{@{}c@{}}35.2 \\(35.2, 35.3)\end{tabular} & \begin{tabular}[c]{@{}c@{}}35.2 \\(35.1, 35.4)\end{tabular} & 8246                      & \begin{tabular}[c]{@{}c@{}}-1750.3 \\(-1758.0, -1738.3)\end{tabular} & \begin{tabular}[c]{@{}c@{}}-1744.7 \\(-1755.6, -1724)\end{tabular} & \begin{tabular}[c]{@{}c@{}}-1709.1 \\(-1741.9, -1606.9)\end{tabular} & 10954                      \\ 
\hline
Lactin   & \begin{tabular}[c]{@{}c@{}}36.9 \\(36.6, 37.3)\end{tabular} & \begin{tabular}[c]{@{}c@{}}38.3 \\(38.1, 38.5)\end{tabular} & \begin{tabular}[c]{@{}c@{}}38.2 \\(37.9, 38.5)\end{tabular} & 7797                      & \begin{tabular}[c]{@{}c@{}}-1693.9 \\(-1703, -1679.6)\end{tabular}   & \begin{tabular}[c]{@{}c@{}}-1764.0 \\(-1780.3, -1738.1)\end{tabular} & \begin{tabular}[c]{@{}c@{}}-1776.2 \\(-1784.9, -1762)\end{tabular}   & 8227                       \\
\hline
\end{tabular}
\begin{tablenotes}
\item{$\dag$} deviance of the model given the data,
\item{$\ddag$} effective sample size.
\end{tablenotes}
\end{threeparttable}
\end{adjustbox}

\begin{table}
\resizebox{\linewidth}{!}{%
\begin{threeparttable}
\captionsetup{width=2\textwidth}
\caption{Model selection criteria for the eight models applied to the \textit{Propylea Coccinellidae} data}
\label{tab:perakis}
\centering
\sisetup{table-number-alignment = center,round-mode = figures,
round-precision = 3}
\begin{tabular}{|c|c|c|c|c|c|c|c|c|c|} 
\hline
                                                                                      & \multicolumn{1}{l|}{} & \multicolumn{1}{l|}{AIC} & \multicolumn{1}{l|}{DIC} & \multicolumn{1}{l|}{LooCV} & \multicolumn{1}{l|}{WAIC} & \multicolumn{1}{l|}{BIC} & \multicolumn{1}{l|}{$log\left(P_{y}IS\right)^\dag$ \, (se)} & \multicolumn{1}{l|}{$log\left(P_{y}PP\right)^\ddag$ \, (se)} & \multicolumn{1}{l|}{$log\left(P_{y}BS\right)$\S \, (se)} \\ 
\hline
\multirow{4}{*}{\begin{tabular}[c]{@{}c@{}}Gaussian\\model \end{tabular}}     & Bieri                 & \textbf{-804.7}          & \textbf{-783.6}          & \textbf{-816.1}            & \textbf{-816.1}           & \textbf{-786.1}          & \textbf{419.9 (19.6)}                           & \textbf{433.4 (18.0)}                           & \textbf{462.4 (9.2)}                             \\ 
\cline{2-10}
                                                                                      & Briere                & -650.5                   & -649.5                   & -658.3                     & -658.3                    & -637.3                   & 299.2 (24.6)                                    & 235.4 (6.7)                                     & 231.4 (1.2)                                      \\ 
\cline{2-10}
                                                                                      & Analytis              & -662.1                   & -637.1                   & -671.4                     & -671.4                    & -646.1                   & 283.7 (24.0)                                    & 312.2 (10.0)                                    & 274.2 (33.7)                                     \\ 
\cline{2-10}
                                                                                      & Lactin                & -789.8                   & -776.8                   & -806.7                     & -806.7                    & -763.3                   & 390.1 (33.5)                                    & 328.1 (23.0)                                    & 334.7 (1.7)                                      \\ 
\hline
\multirow{4}{*}{\begin{tabular}[c]{@{}c@{}}Inverse\\ Gamma\\model \end{tabular}} & Bieri                 & -716.1                   & -719.3                   & -719.3                     & -719.3                    & -702.8                   & 334.5 (25.9)                                    & 342.9 (31.5)                                    & 333.2 (5.0)                                      \\ 
\cline{2-10}
                                                                                      & Briere                & -560.8                   & -561.4                   & -561.4                     & -561.4                    & -550.2                   & 268.2 (23.2)                                    & 265.7 (4.6)                                     & 251.7 (1.9)                                      \\ 
\cline{2-10}
                                                                                      & Analytis              & -734.7                   & -732.4                   & -732.4                     & -732.4                    & -718.8                   & 340.8 (26.7)                                    & 371.3 (14.5)                                    & 335 (31.2)                                       \\ 
\cline{2-10}
                                                                                      & Lactin                & -698.9                   & -699.1                   & -699.1                     & -699.1                    & -685.6                   & 329.2 (19.5)                                    & 329.3 (11.8)                                    & 313.7 (18.8)                                     \\
\hline
\end{tabular}
\begin{tablenotes}
\item{${\dag}$} $log\left(P_{y}IS\right)$ denotes the logarithm of estimated marginal likelihood via Importance sampling,
\item{$\ddag$} $log\left(P_{y}PP\right)$ denotes the logarithm of estimated marginal likelihood via Power posterior,
\item{$\S$} $log\left(P_{y}BS\right)$\S denotes the logarithm of estimated marginal likelihood via Bridge sampling.
\end{tablenotes}
\end{threeparttable}}
\end{table}

\section{Posterior prediction graphs for both datasets}

\begin{figure}[hbt!]
\begin{center}
\includegraphics[height=6cm,width=\linewidth]{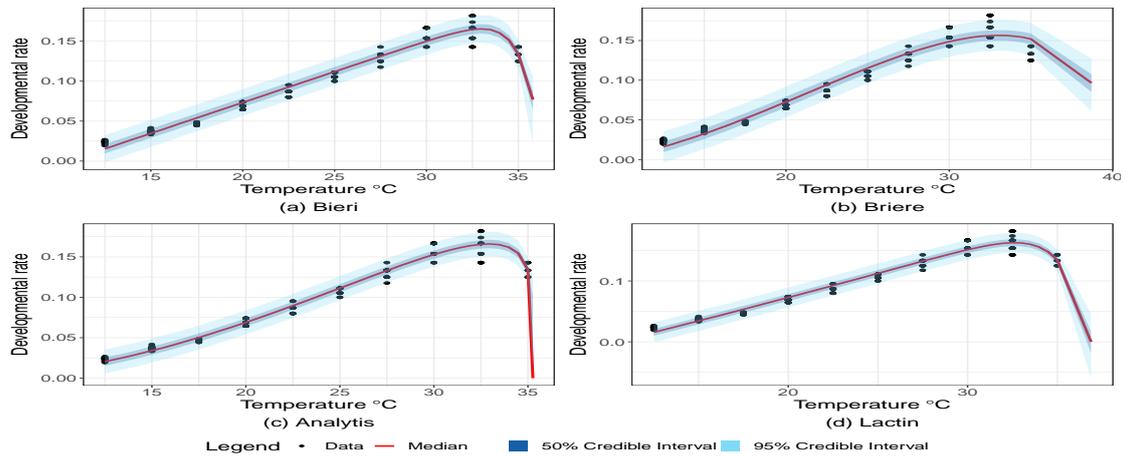}
    \caption{Predicted posteriors versus \textit{Tetranychus urticae} data using Gaussian distribution}%
    
\label{fig:MFL3}%
    \end{center}
\end{figure}

\begin{figure}[hbt!]
\begin{center}
\includegraphics[height=6cm,width=\linewidth]{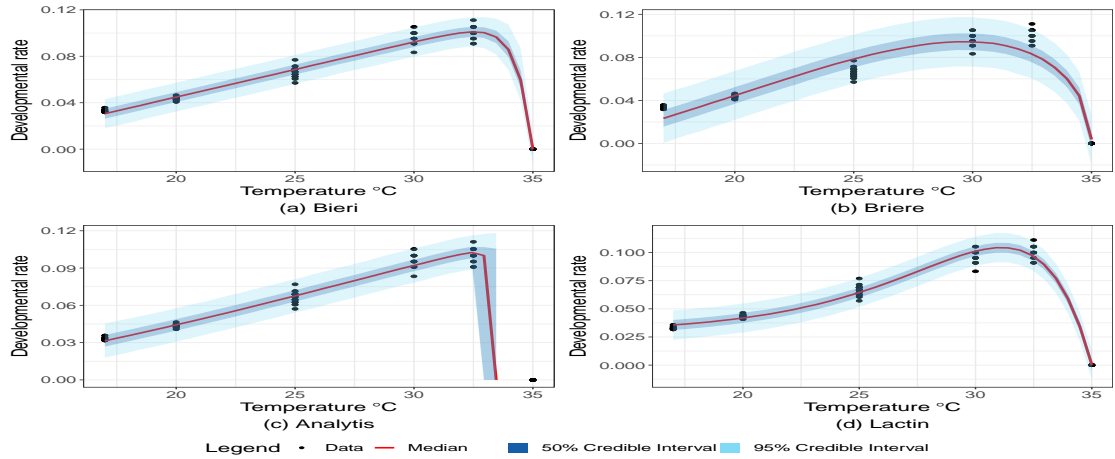}
    \caption{Predicted posteriors versus Propylea Coccinellidae data using Gaussian distribution}%
    
\label{fig:MFL1}%
    \end{center}
\end{figure}
\newpage

\begin{figure}[hbt!]
\begin{center}

     \includegraphics[height=6cm,width=\linewidth]{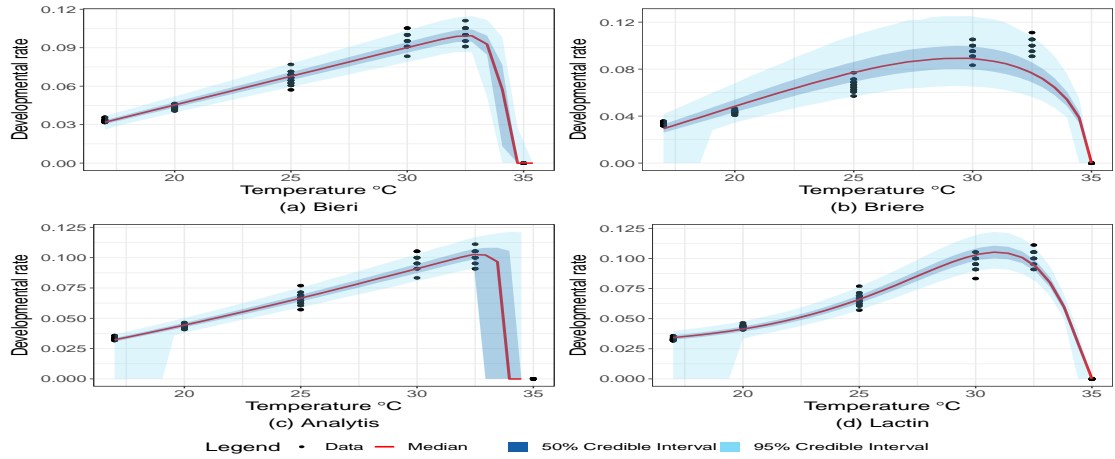}
    \caption{Predicted posteriors versus Propylea Coccinellidae data using Zero Inflated Inverse Gamma distribution}%
    \label{fig:MFL2}%
    \end{center}

\end{figure}

\section{Bayesian inference}
\label{bayesian_details}
In the current section, we provide details of statistical model specifications in the Bayesian framework. Specifically, we provide a brief description of the most important features of Stan’s implementation of HMC and VB so the reader can get familiar with the tools that Stan is based on. We then provide model selection and model averaging techniques in order to compare the different ecological models and to explore and interpret the parameters of interest combining  predictions from all the four of them. 

\subsection{HMC and VB techniques}
The HMC method is a Monte Carlo technique that uses Hamiltonian dynamics in order not only to explore efficiently the target distribution but also to propose distant samples in the parameter space that do not exclusively depend on the current state of the Markov chain like considered in previous MCMC methodology \citep{neal1901mcmc}. In this way, many performance challenges are tackled like either the low convergence due to the fact that the parameter space with high posterior support is not reached or the poor exploration of the target distribution due to its multi-modality or its shape irregularities. The existence of Hamiltonian dynamics in the system of the joint density mass function allows the preservation of volume and hence adequate trajectories can be used to define complex mappings of the parameter state space without the need to account for cumbersome Jacobian calculations \citep{barber2003biocontrol}. Thus, by carefully designing automated trajectory realizations in the Hamiltonian dynamics system, the Stan team managed to create an augmented software called STAN \citep{carpenter2017stan}, which materializes HMC sampling for the parameters of interest.

Moreover, independently, Automatic Differentiation Variational Inference (ADVI) technique is referred to the machine learning field \citep{blei2017variational}. The latter is a VB method and posterior target distributions are approximated by choosing the closest distribution to a parametric family of tractable distributions like the exponential family via optimization. In order to achieve this point-wise estimations of the parameters of the family distribution are estimated so that the Kullback–Leibler `KL' divergence function is minimized. Specifically, since the KL divergence is intractable the Evidence Lower Bound is maximized instead \citep{blei2017variational}.

\subsection{Model selection and model averaging}
In case there are $m$ models $(\mathcal{M}_1,\ldots,\mathcal{M}_m)$ under consideration, the posterior probability of the suitability of the $i^{th}$ model given the data ${y}$, is given by
\begin{equation}\label{modposterior}
p(\mathcal{M}_i|{y})=\frac{p({y}|\mathcal{M}_i)p(\mathcal{M}_i)}{\sum_{k=1}^{m}p({y}|\mathcal{M}_k)p(\mathcal{M}_k)}
\end{equation}
where $p(\mathcal{M}_i)$ expresses the prior belief for the $i^{th}$ model, while the $p({y}|\mathcal{M}_i)$ is the model evidence also called `marginal likelihood' and it can be interpreted as the likelihood over the space of models, marginalizing out the parameters of the $i^{th}$ model. The ratio of the marginal likelihoods between two models $\frac{p({y}|\mathcal{M}_i)}{p({y}|\mathcal{M}_j)}$ is called Bayes factor and is the posterior odds of the null hypothesis that the $i_{th}$ model fits better the data than the $j_{th}$ model does when the prior probability of the null is one-half \citep{kass1995bayes}. The Bayes factor is used in order to give evidence for the most probable model given the data, when comparing two alternative models \citep{kass1995bayes}.

In the case of Bayesian model averaging, the model selection uncertainty is taking into account in statistical inference. The joint posterior $p(\mathcal{M}_i,{\theta}_i | {y})$ of the $i^{th}$ model with vector of parameters $\theta_i$, using the Baye's rule, is proportional to the product of the likelihood of the $i^{th}$ model times the prior distribution of the parameters $p({\theta})$ times the prior distribution $p(\mathcal{M}_i)$ (that expresses our uncertainty of the $i^{th}$ model)
\begin{equation}\label{jointp}
p(\mathcal{M}_i,{\theta_i} | {y}) \propto p({y}|{\theta}_i,\mathcal{M}_i) \cdot p({\theta}) \cdot p(\mathcal{M}_i).  
\end{equation}
The uncertainty of the $i^{th}$ model given the data can then be re-expressed via the posterior probability $p(\mathcal{M}_i|{y})$ defined by the ratio in (\ref{modposterior}), in case of existence of multiple models. The posteriors of the models can be thought as weights that are critical to the Bayesian model averaging as they can be used to extract useful weighted statistics from the data distribution while at the same time taking into account model uncertainties.
Estimation of model parameters and model uncertainties can be achieved either by directly sampling from the joint posterior (\ref{jointp}) or by approximating the marginal likelihood of each model independently and, accordingly, by controlling the outcomes with a view to formulating proper weights and proceeding with the calculation of the averaged statistics. For the former case, techniques like the reversible jump MCMC \citep{green1995reversible, george1997approaches} and variable selection samples \citep{carlin1995bayesian, kuo1998variable, dellaportas2000bayesian} are used. On the other hand, for the later case, techniques of marginal likelihood approximations via thermodynamic integration \citep{friel2008marginal}, bridge sampling \citep{meng1996simulating}, importance sampling \citep{perrakis2014use} or via information- criteria perspective like in \citep{kass1995bayes} are used.    

\subsubsection{Information criteria}
The criteria used in the current work are the Akaike information criterion `AIC', the Bayesian information criterion `BIC', the Deviance information criterion `DIC', the Watanabe–Akaike information criterion `WAIC' and the  Leave-one-out cross-validation criterion `Loocv'. Briefly, these criteria provide an approximation of the expected log predictive density for new-coming data while correcting bias from data usage. In particular AIC  \citep{Akaike1100705} is defined as the difference
\begin{equation*}\label{AIC}
AIC(\mathcal{M}_i)=-2\log(y|\widehat{\theta _i})+2k_{i}
\end{equation*}
where $\widehat{\theta_i}$ is the Maximum Likelihood estimate `MLE' of the $k_i$ parameters of the $i^{th}$ model. Similarly, BIC
\citep{schwarz1978estimating} is defined as the difference
\begin{equation*}\label{BIC}
BIC( \mathcal{M}_i)=-2\log(y| \widehat{ \theta_i})+k_{i} \cdot \log(n)
\end{equation*}
where n is the sample size.
In addition, DIC \citep{spiegelhalter2002bayesian} is defined as the following difference:
\begin{equation*}\label{DIC}
DIC(\mathcal{M}_i)=-2\log p(y|\widehat{\theta_{i}})+2p_{DIC}
\end{equation*}
where $\widehat{\theta_{i}}$ is the posterior mean of the parameters of the $i^{th}$ model, whereas $p_{DIC}$ is the effective number of parameters and it is evaluated following \citep{spiegelhalter2002bayesian, gelman2013bayesian} by either
\begin{equation*}\label{pdic1}
p_{DIC_1}=E_{\theta|y}\{-2\log {p}\left(y|\theta \right)\}+2\log \left\{{p} \left( y|\widehat{\theta_{i}} \right) \right\},
\end{equation*}
or 
\begin{equation*}\label{pdic2}
p_{DIC_2}=\frac{Var_{\theta|y}\left\{\log p(y|\theta)\right\}}{2},
\end{equation*}
where $E_{\theta|y}\left\{\log p(y|\theta)\right\}$ is an expectation over the posterior density of $\theta$, whereas $Var_{\theta|y}\left\{\log p(y|\widehat{\theta})\right\}$ is the variance of the log posterior density of the observed data y, over the posterior density of $\theta$.
Furthermore, WAIC \citep{watanabe2010asymptotic} is defined as the following difference:
\begin{equation*}\label{WAIC}
WAIC(\mathcal{M}_i)=-2\sum_{j=1}^{N}  \log E_{\theta_{i}}\left\{ p(y_j|\theta_{i}) \right \}+2p_{WAIC}
\end{equation*}
where $E_{\theta_{i}}\left\{ p(y_j|\theta_{i}) \right \}$ is the expectation of the  probability at $y_{j}$ data point over the posterior distribution of the parameters of the $i^{th}$ model, whereas $p_{WAIC}$ is the effective number of parameters and it is evaluated following \citep{gelman2013bayesian} by either

\begin{equation*}\label{pwaic1}
p_{{WAIC}_{1}} = 2\sum_{j=1}^{N}\left[\log E_{p(\theta|y)}\left\{p(y_{j}|\theta)\right\}-E_{p(\theta|y)}\left\{\log p(y_{j}|\theta)  \right \}\right]
\end{equation*}

or 
\begin{equation*}\label{pwaic2}
p_{{WAIC}_{2}}=\sum_{j=1}^{N}var_{p(\theta|y)}\left \{ \log  p(y_{j}|\theta)  \right \}
\end{equation*}
where $E_{p(\theta|y)}\left\{\log p(y_{j}|\theta)  \right \}$ is the expectation over the logarithm of the posterior density of $\theta$ at $y_{j}$ data point, whereas $var_{p(\theta|y)}\left \{ \log  p(y_{j}|\theta)  \right \}$ is the variance of the log posterior density of the observed data $y_{j}$, over the posterior density of $\theta$.

Furthermore, LooCV \citep{gelman2013bayesian} is defined as the following difference:
\begin{equation*}\label{LooCV}
LooCV(\mathcal{M}_i)=- 2\sum_{j=1}^{N}\log E_{\theta_{i}^{-j}}\left\{ p(y_j|\theta_{i}^{-j}) \right \}-2\beta_{LooCV}
\end{equation*}
where $E_{\theta_{i}^{-j}}\left\{ p(y_j|\theta_{i}^{-j}) \right \}$ is the expectation of the  probability at $y_{j}$ data point over the posterior distribution of the parameters of the $i^{th}$ model. The posterior distribution $p({\theta_{i}^{-j}}|y_{-j})$ is sampled considering a partition of the data, leaving one data value ($y_{j}$) out of the original sample. The $\beta_{LooCV}$ is a bias correction of the measure and it is evaluated following \citep{gelman2013bayesian} by

\begin{equation*}\label{bloocv}
\beta_{Loocv} = \sum_{j=1}^{N}  \log E_{\theta_{i}}\left\{ p(y_j|\theta_{i}) \right \}-\frac{1}{N}\sum_{\kappa =1}^{N}\sum_{j=1}^{N}\log E_{\theta_{i}^{-\kappa }}\left\{ p(y_j|\theta_{i}^{-\kappa }) \right \}
\end{equation*}
where $E_{\theta_{i}^{-\kappa }}\left\{ p(y_j|\theta_{i}^{-\kappa }) \right \}$ is the expectation of the  probability at $y_{j}$ data point over the posterior distribution of the parameters of the $i^{th}$ model leaving out the $\kappa^{th}$ observation.

\subsubsection{Marginal likelihood estimation techniques}
The marginal likelihood can be viewed as a normalizing constant $z_i=p({y}|\mathcal{M}_i)$ of the density $q({\theta_i}|{y})=p({y}|{\theta_i})\cdot p({\theta_i})$ within the $i^{th}$ ecological model that includes parameters $\theta_i$. In the general scheme of comparing the two densities $q_0$ and $q_1$ of interest, as in the case of the Bayes factor of two models or in the case model's prior and posterior, a general path from $q_0$ to $q_1$ can be created according to \citep{gelman1998simulating} using a class of densities $p(\theta_i|{y},t)$ on the same space indexed by the continuous auxiliary variable say $t \in [0, 1]$. A key formula that links the corresponding normalizing constant $z(t)$ and the unormalized density $q(\theta_i|{y},t)$ that correspond to the sampling distribution $p(\theta_i|{y},t)$ is given by:  
\begin{equation}\label{sampl_scheme}
\frac{d}{dt} \log z(t)=\int {\frac{1}{z(t)}\frac{d}{dt}q({\theta_i}|{y},t)p({\theta_i}|t)d{\theta_i}} =E_{t} \left \{ \frac{d}{dt}\log{q({\theta_i}|{y},t)} \right \},
\end{equation}
where the expectation is with respect to the sampling distribution $p(\theta_i|{y},t)$. 

In addition, another key formula of estimating a ratio of normalizing constants has been of great interest such as in computing likelihood ratios in hypothesis testing or in computational physics in estimating free energy differences, or in computing the Bayes factor in Bayesian framework \citep{meng1996simulating}. The general formula is as follows:
\begin{equation}\label{bridge_ratio}
\frac{z_1}{z_0}=\frac{p({y}|\theta,t=1)}{p({y}|\theta,t=0)}=\frac{E_0 \left \{ h(\theta) \cdot q(\theta|{y},t=1)\right \} } {E_1 \left \{ h(\theta) \cdot q(\theta|{y},t=0) \right \} }
\end{equation}
where $E_0$ and $E_1$ expectations are with respect to posterior distribution densities $p(\theta|{y},t=0)$ and $p(\theta|{y},t=1)$ respectively, whereas the bridge function $h(\theta)$ is defined and overlapped by the common support of the former densities.

Using general formulas (\ref{sampl_scheme}) and (\ref{bridge_ratio}), several marginal probability evaluation schemes of the $i^{th}$ model are derived \citep{gelman1998simulating}. The power posterior sampling \citep{friel2008marginal}, the importance sampling \citep{perrakis2014use} and the bridge sampling \citep{meng1996simulating, overstall2010default} techniques are used for the current work.

In the power posterior case, formula (\ref{sampl_scheme}) is integrated with respect to variable t and $q(\theta_i|{y},t)$ is substituted with density $p({\theta_i}|y)^{t} p(\theta_i)$. The marginal likelihood $z_i=p({y}|\mathcal{M}_i)$ is derived from logarithmic scale by the equation:
\begin{equation}\label{power_post}
\log\left \{p({y}|\mathcal{M}_i)\right \}=\int_{0}^{1}E_{\theta_{i}|y,t} \left \{ \log{p({y}|\theta_{i})} \right \}dt
\end{equation}
where expectation $E_{\theta_{i}|y,t}$ is taken with respect to the density $p({\theta_i}|y)^{t} p(\theta_i)$ which is defined as the power posterior at temperature t \citep{friel2008marginal}. 

Additionally, the standard error ${se}_i$ for the ith model estimator (\ref{power_post}), as shown in section \ref{pp_variance} of the appendix is approximated by:

\begin{equation*}\label{power_post_se}
\hat{{se}_i}=\sqrt{\frac{\left (t_2-t_1 \right)^2}{2}s_1^2+\sum_{k=2}^{n-1}\frac{\left (t_{k}-t_{k-1} \right)^2}{2}s_k^2+\frac{\left (t_n-t_{n-1} \right)^2}{2}s_n^2 },
\end{equation*}
where $t_k$ is the time after discretization $0=t_0<t_1<t_k<t_n=1$ and $s_k$ is the standard error of the corresponding estimation $\log\left \{p({y}|\mathcal{M}_i)\right \}$ given in (\ref{power_post}).

In the case of importance sampling, the marginal likelihood is assessed by introducing the proper density function \textit{g}. After sampling from the proposed density function \textit{g}, the marginal likelihood is calculated as with respect to \textit{g} as:

\begin{equation*}\label{marginal_imp0}
p({y}|\mathcal{M}_i)=E_{g} \left \{ \frac{q(\theta_i|{y})}{g(\theta_i)}\right \}
\end{equation*}

Following \citep{perrakis2014use}, we use the density $q(\theta_i|{y})$ equal to $p({y}|\theta_i,\phi_i)\cdot p(\theta_i,\phi_i)$ and the auxiliary importance function \textit{g} used is as follows: 
\begin{equation}\label{marginal_sugg}
g(\theta_i)=g(\theta_i,\phi_i)=p(\theta_i|{y})p(\phi_i|{y}),
\end{equation}
where $(\theta_i,\phi_i)$ are the parameters of the $i^{th}$ model  divided into two blocks $\theta_i$ and $\phi_i$ which may or may not be independent. The right hand side of (\ref{marginal_sugg}) is the product of the marginal posterior distributions of the block. Thus, the marginal probability which gives the target value is given as follows:

\begin{equation}\label{marginal_imp1}
p({y}|\mathcal{M}_i)=\iint_{}^{}\frac{p({y},\theta_i,\phi_i)}{g(\theta_i,\phi_i)}g(\theta_i,\phi_i)d(\theta_i,\phi_i)=E_{g} \left \{ \frac{p({y},\theta_i,\phi_i)}{g(\theta_i,\phi_i)}\right \}
\end{equation}
The standard error ${se}_i$ of (\ref{marginal_imp1}) as shown in section \ref{IS_variance} of the appendix is:

\begin{equation*}\label{marginal_imp1_se}
{\hat{{se}_i}}=\sqrt{\frac{1}{K}\sum_{j=1}^{K}\left\{\frac{p({y}|\theta^j,\phi^j)\cdot p(\theta^j,\phi^j)}{g\left(\theta^j\right)}-{\hat{z}}_i\right\}^2},
\end{equation*}
where $\widehat{z}_i$ is the estimation of the corresponding marginal probability (in the same form of (\ref{marginal_imp1})), while $(\theta^j,\phi^j)$ are draws $j=1,2,\dots,K$ from the importance function in (\ref{marginal_sugg}).\\

Additionally, using an alternative version of (\ref{bridge_ratio}) in \citep{meng1996simulating, fruhwirth2004estimating,overstall2010default} the marginal likelihood of a single model is evaluated using bridge sampling by the formula:
\begin{equation}\label{bridge_ratio2}
z_i=p({y}|\mathcal{M}_i)=\frac{E_g \left \{ h(\theta_i) \cdot q(\theta_i|{y})\right \} } {E_p \left \{ h(\theta_i) \cdot g(\theta_i)\right \} },
\end{equation}
where $E_g$ and $E_p$ are the expectations with respect to $g(\theta_i)$ a so-called proposal distribution and to $p(\theta_i|{y})$ the $i^{th}$ model posterior distribution respectively. 

The bridge function $h(\theta_i)$ is selected to minimize the relative mean-squared error of (\ref{bridge_ratio}).
Following \citep{meng1996simulating} the bridge function is specified by:
\begin{equation}\label{bridge_optimal}
h(\theta_i)= \mathcal{C}\cdot\frac{1} {s_{1} \cdot q(\theta_i|{y})+ s_{2} \cdot p(y) \cdot g(\theta_i)},
\end{equation}
where $s_{1}=\frac{N_{1}}{N_{1}+N_{2}} $, $s_{2}=\frac{N_{2}}{N_{1}+N_{2}} $ and $\mathcal{C}$ is a constant. $N_{1}$ is the sample size from the posterior and $N_{2}$ is the sample size from $g(\theta_i)$. 

The optimal bridge function in (\ref{bridge_optimal}) includes the marginal likelihood under-assessment so that it cannot be evaluated directly. For this purpose the iterative method suggested by \citep{meng1996simulating} and applied in \citep{bridgesampling2020} in R software \citep{rteam} is used. 
The alternatives used in place of distribution $g$ is either  a multivariate normal distribution with mean vector and covariance matrix that match the respective posterior samples quantities or a standard multivariate normal distribution in combination with a warped posterior distribution of which the first three moments correspond to \citep{bridgesampling2020}.

Moreover, following \citep{fruhwirth2004estimating} the relative mean square error ${RE}^2_i=\frac{E\left\{\widehat{z_i}-z_i\right\}^2}{z^2_i}$ of (\ref{bridge_ratio2}) is evaluated by the formula:
\begin{equation}\label{bridge_ratio2_rmse}
\widehat{RE}^2_i={\frac{1}{N_2} \frac{V_{g} \left \{ f_1(\theta_i) \right \} }{E_{g}^2 \left \{f_1(\theta_i) \right \} }+ \frac{\rho_{f_2}(0)}{N_1}\frac{V_{p} \left \{f_2(\theta_i) \right \}}{E_{p}^2 \left \{f_2(\theta_i) \right \}}},
\end{equation}
where $f_1(\theta_i)=\frac{q(\theta_i|{y})} {s_{1} \cdot q(\theta_i|{y})+ s_{2} \cdot g(\theta_i)}$, $f_2(\theta_i)=\frac{g(\theta_i)} {s_{1} \cdot q(\theta_i|{y})+ s_{2} \cdot g(\theta_i)}$, \\ $V_{g}(f_1(\theta_i))=\int_{}^{} \left \{f_1(\theta_i)-E(f_1(\theta_i)) \right \}^2 g(\theta_i)d\theta$ is the variance of $f_1(\theta_i)$ with respect to $g(\theta_i)$. The term $\rho_{f_2}(0)$ in (\ref{bridge_ratio2_rmse}) corresponds to the normalized spectral density of the auto-correlated process $f_2(\theta_i)$ at the frequency 0.

Following \citep{bridgesampling2020} the square root of ${RE}^2$ can be interpreted as coefficient of variation provided that the bridge sampling estimator $\widehat{z}_i$ is unbiased. Then the stantard error ${se}_i$ of the bridge estimator is evaluated by the product $\widehat{se}_i=\widehat{RE} \cdot E(\widehat{z}_i)$

\subsubsection{BMA weights}
\label{BMA_theory}
We can derive a weighted prediction $\tilde{y}$ over the m different models $\mathcal{M}_1,\mathcal{M}_2, \dots  \mathcal{M}_m$ predictions $\hat{y}_{1},\hat{y}_{2},\dots \hat{y}_{m}$ by imposing appropriate weights $w_1,w_2, \dots  w_m$. 
\begin{equation*}\label{prediction_weights}
\tilde{y}=\sum_{i=1}^{m}{\hat{y}_{i} \cdot w_i}  \quad \textrm{and} \quad  \sum_{i=1}^{m}{w_i}=1 .
\end{equation*}
In the Bayesian framework, model weights definition is straightforward. The model weights used are the posterior model weights $w_{i}=p(\mathcal{M}_i | {y})$ given in (\ref{modposterior}) that represent the relative probability of each model given the data. So a major challenge is to estimate these Bayesian weights. Except for using the marginal likelihood estimations mentioned in previous section, we also investigate approximations of the weights by using the BIC for each model. In particular, model weights can be estimated through the following equations \citep{kass1995bayes,buckland1997model}:
\begin{equation}\label{IC_weights}
\begin{aligned}
\\{w}_{i}=\frac{e^{-0.5 \cdot \left (BIC(\mathcal{M}_i) \right ) }}{
\sum_{j=1}^{m}{e^{-0.5 \cdot \left ( BIC(\mathcal{M}_j) \right )}}} .
\end{aligned}
\end{equation}
Instead of BIC, the AIC, DIC, WAIC and LooCV  are also used in (\ref{IC_weights}).
We investigate both approaches in insect observed rates and compare the results taking into account model complexity, data scarcity and Biological interpretation. 

\section{Power posterior for Gaussian and Inverse Gamma distribution}
\label{pp_distribution}
Following \citep{friel2008marginal} the power posterior in (\ref{power_post}) includes the likelihood raised to the power of t $p({\theta_i}|y)^{t}$. 

In the the Gaussian case, the likelihood involved becomes:
\begin{equation*}
\begin{aligned}
p(y|{\theta})^{t}=p^{t}\left( y|\mu,\sigma^2\right)=\frac{e^{-\frac{1}{2}\left(\sqrt t\cdot\frac{y-\mu}{\sigma}\right)^2}}{\sqrt{2\pi\sigma^2}}=p\left(y|\mu,\frac{\sigma^2}{t}\right)\cdot{\frac{\sqrt{t}}{t}} .
\end{aligned}
\end{equation*}
Inserting the current ecological model $r \left ( T;\theta \right )$, the log of the power posterior is given by:
\begin{equation*}
\begin{aligned}
 log\left (p(y|{\theta})^{t}  \right )=-\frac{1}{2}\log{\left(2\pi\sigma^2\right)}-\frac{1}{2}\left(\sqrt t\cdot\frac{y-r\left(T;\theta \right)}{\sigma}\right)^2 .
\end{aligned}
\end{equation*}

In the Inverse Gamma case, the likelihood involved becomes:
\begin{equation*}
\begin{aligned}
p(y|{\theta})^{t}=p(y|{\alpha, \beta})^{t}=\frac{\beta ^{\alpha \cdot t}}{\Gamma^{t} \left ( \alpha  \right )}y^{-\alpha\cdot t -t}exp\left ( -\frac{\beta \cdot t}{y} \right )=\\
\frac{\left (\beta\cdot t  \right ) ^{\alpha \cdot t}}{\Gamma \left ( \alpha  \cdot t \right ) }\frac{\Gamma \left ( \alpha  \cdot t \right )}{\Gamma^{t} \left ( \alpha  \right )\cdot (t) ^{\alpha \cdot t}}y^{-\alpha\cdot t -1}y^{1 -t}exp\left ( -\frac{\beta \cdot t}{y} \right )=\\
In G\left ( \alpha \cdot t, \beta \cdot t\right )\cdot \frac{\Gamma \left ( \alpha  \cdot t \right )}{\Gamma^{t} \left ( \alpha  \right )\cdot (t) ^{\alpha \cdot t}}\cdot y^{1 -t} .
\end{aligned}
\end{equation*}
Inserting the current ecological model $r \left ( T;\theta \right )$, the log of the power posterior is given by:
\begin{equation*}
\begin{aligned}
 log\left (p(y|{\theta})^{t}  \right )={\alpha \cdot t} \cdot \left ( log\left( \alpha -1 \right )+ log\left ( r \left ( T;\theta \right ) \right )  \right )-\\
t\cdot log\left ( \Gamma \left ( \alpha  \right ) \right )-t\cdot \left ( \alpha +1 \right )\cdot log\left ( y  \right )-\frac{ \left ( \alpha -1 \right )\cdot \left (r \left ( T;\theta \right ) \right ) \cdot t}{y} .
\end{aligned}
\end{equation*}
\section{Estimation of Variance of Power posterior method}
\label{pp_variance}
\begin{equation*}
    \begin{aligned}
\hat{\sigma_y}^2=Var\left\{log P\left(y|\theta\right)\right\}= Var\left\{\int_{0}^{1}{ E\left\{{log}P{\left(y |\theta\right)}\right\}dt} \right\}\simeq\\
\sum_{i=1}^{n-1}{ \frac{1}{4}Var\left\{E_{\theta |y,t_i} \left\{{log}P{\left(y |\theta\right)}\right\}+E_{\theta |y,t_i+1} \left\{{log}P{\left(y |\theta\right)}\right\} \right\}\left(t_{i+1}-t_i\right)^2}\simeq\\
\sum_{i=1}^{n-1}{\left[\frac{{{sd}^2}_{\theta |y,t_i}}{2}\right] \cdot\left(t_{i+1}-t_i\right)^2}
    \end{aligned}
\end{equation*}
where ${{sd}}_{\theta |y,t_i}$ is the std error estimated at the $t_i$ temperature
\section{Estimation of Variance of Importance sampling method}
\label{IS_variance}
The marginal likelihood estimate for the ith model is given by
$m=p\left( y|M_i\right)=E_g\left\{\frac{q\left( \theta_i|y\right)}{g\left(\theta_i\right)}\right\}=\int\frac{q\left(\theta_i|y\right)}{g\left(\theta_i\right)}g\left(\theta_i\right)d\theta_i$ provided $g\left(\theta
-i\right)>0$ whenever $q\left( \theta_i|y\right)\neq0$
where the density $q\left(\theta_i\middle|\ y\right)$ is equal to $p\left(y\middle|\theta_i,\phi_i\right)\cdot p\left(\theta_i,\phi_i\right)$ and the auxiliary importance function $g$ used is the following: 
$g\left(\theta_i\right)=g\left(\theta_i,\phi_i\right)=p\left(\theta_i\middle|\ y\right)p\left(\phi_i\middle|\ y\right)$ \\
Removing the ith model index we can evaluate the marginal likelihood via MC integration which gives the formula below:
\begin{equation*}
{\hat{z}}_y=\hat{p}\left(y\right)=\frac{1}{M}\sum_{J=1}^{M}\frac{q\left(\theta^j|y\right)}{g\left(\theta^j\right)}
\end{equation*}

where $\theta^j,j=1,2,3,\ldots,M$ are obtained from density: $g\left(\theta_i\right)$.\\
\begin{equation*}
E_g\left({\hat{z}}_y\right)=\frac{1}{M}\sum_{j=1}^{M}{E_g\left\{\frac{q\left(\theta^j|y\right)}{g\left(\theta^j\right)}\right\}}=p\left(y\right)=m .
\end{equation*}
Assuming that  $Cov\left\{\frac{q\left( \theta^j|y\right)}{g\left(\theta^j\right)},\frac{q\left( \theta^i|y\right)}{g\left(\theta^i\right)}\right\}=0,\ for\ i\neq\ j$ \\

$V_g\left({\hat{z}}_y\right)=V_g\left(\frac{1}{M}\sum_{j=1}^{M}\left\{\frac{q\left( \theta^j|y\right)}{g\left(\theta^j\right)}\right\}\right)=\frac{1}{M^2}\sum_{j=1}^{M}{V_g\left\{\frac{q\left( \theta^j|y\right)}{g\left(\theta^j\right)}\right\}+}\frac{1}{M^2}\sum_{k,j=1}^{M}{{Cov}_g\left\{\frac{q\left( \theta^k|y\right)}{g\left(\theta^k\right)},\frac{q\left( \theta^j|y\right)}{g\left(\theta^j\right)}\right\}=}$\\
$\frac{1}{M^2}\sum_{j=1}^{M}{V_g\left\{\frac{q\left( \theta^j|y\right)}{g\left(\theta^j\right)}\right\}}= 
\frac{1}{M}V_g\left\{\frac{q\left( \theta|y\right)}{g\left(\theta\right)}\right\}=\frac{1}{M}\left\{{E_g\left\{\frac{q\left( \theta|y\right)}{g\left(\theta\right)}\right\}}^2-{E_g}^2\left\{\frac{q\left( \theta|y\right)}{g\left(\theta\right)}\right\}\right\}=$\\
$\frac{1}{M}\int_{G}{\left\{\frac{q\left( \theta|y\right)}{g\left(\theta\right)}\right\}^2g\left(\theta\right)d\theta}-\frac{1}{M}m^2\Rightarrow$ \\
Also, using properties of expectation and the fact that $g\left(\theta\right)$ is proper \\
$V_g\left({\hat{z}}_y\right)=\frac{1}{M}\int_{G}{\left\{\frac{q\left( \theta|y\right)-m \cdot g\left(\theta\right)}{g\left(\theta\right)}\right\}^2g\left(\theta\right)d\theta}=\frac{\sigma_y^2}{M}$
The latest can be used to estimate the std error of the estimator ${\hat{z}}_y$ via the formula:
\begin{equation*}
{\hat{\sigma}}_y=\sqrt{\frac{1}{K}\sum_{j=1}^{K}\left\{\frac{q\left( \theta^j|y\right)}{g\left(\theta^j\right)}-{\hat{z}}_y\right\}^2}
\end{equation*}

where $\theta^j,j=1,2,3,\ldots,K$ are obtained from density $g\left(\theta_i\right)$.

\end{document}